\documentclass[prd,aps,floats,floatfix,eqsecnum,nofootinbib]{revtex4}
\usepackage{amsmath,amssymb,verbatim,epsfig,graphicx,rotating}
\usepackage{psfrag}
\newcommand{\be}{\begin{equation}}
\newcommand{\ee}{\end{equation}}
\newcommand{\bea}{\begin{eqnarray}}
\newcommand{\eea}{\end{eqnarray}}
\newcommand{\vk}{\vec{k}}
\newcommand{\vp}{\vec{p}}
\newcommand{\td}{ \delta}
\begin{document}
\title{The dark matter transfer function:
free streaming, particle statistics \\ and memory of gravitational
clustering.}
\author{\bf D. Boyanovsky $^{(a,b,c)}$}
\email{boyan@pitt.edu}
\author{\bf H. J. de Vega$^{(b,c)}$} \email{devega@lpthe.jussieu.fr}
\author{\bf N. G. Sanchez$^{(c)}$} \email{Norma.Sanchez@obspm.fr}
\affiliation{$^{(a)}$ Department of Physics and Astronomy,
University of Pittsburgh, Pittsburgh, Pennsylvania 15260,
USA.\\
$^{(b)}$ LPTHE, Laboratoire Associ\'e au CNRS UMR 7589,\\
Universit\'e Pierre et Marie Curie (Paris VI) et Denis Diderot (Paris VII),\\
Tour 24, 5 \`eme. \'etage, 4, Place Jussieu, 75252 Paris, Cedex 05, France.\\
$^{(c)}$ Observatoire de Paris, LERMA, Laboratoire Associ\'e au CNRS
UMR 8112,  \\61, Avenue de l'Observatoire, 75014 Paris, France.}

\date{\today}

\begin{abstract}
The transfer function $T(k)$ of dark matter (DM) perturbations
during matter domination is obtained by solving  the linearized collisionless
Boltzmann-Vlasov equation.  We provide an \emph{exact} expression
for $T(k)$ for \emph{arbitrary distribution functions of decoupled
particles and initial conditions}, which can be systematically expanded
in a Fredholm series. An exhaustive numerical study of thermal
relics for different initial conditions  reveals that the first
\emph{two} terms in the expansion of $T(k)$ provide a remarkably
accurate and simple approximation valid on all scales of
cosmological relevance for structure formation in the linear regime.
The natural scale of suppression is the free streaming wavevector at
matter-radiation equality, $   k_{fs}(t_{eq}) =
\left[{4\pi\rho_{0M}}/{[\langle \vec{V}^2\rangle\,(1+z_{eq})]}
\right]^\frac12 $. An important ingredient is a non-local kernel
determined by the distribution functions of the decoupled particles
which describes the \emph{memory of the initial conditions and
gravitational clustering} and yields a correction to the fluid
description. This correction is negligible at large scales $k \ll
k_{fs}(t_{eq})$ but it becomes
important at small scales $k \geq k_{fs}(t_{eq})$. Distribution
functions that favor the small momentum region yield longer-range
memory kernels and lead to an \emph{enhancement of power at small
scales} $ k > k_{fs}(t_{eq}) $. Fermi-Dirac and Bose-Einstein
statistics lead to long-range memory kernels, with longer range for
bosons, both resulting in enhancement of $T(k)$ at small scales.
 For DM thermal relics that decoupled while ultrarelativistic we find
$ k_{fs}(t_{eq}) \simeq  0.003 \; (g_d/2)^\frac13 \;
(m/\mathrm{keV}) \; [\mathrm{kpc}]^{-1} $, where $ g_d $ is the
number of degrees of freedom at decoupling.   For WIMPS we obtain $
k_{fs}(t_{eq}) = 5.88 \; (g_d/2 \; )^\frac13 \; (m/100\,\mathrm{GeV}
)^\frac12 \; (T_d/10 \,\mathrm{MeV})^\frac12 \; [\mathrm{pc}]^{-1}
$. For $k\ll k_{fs}(t_{eq})$, $T(k) \sim 1-
\mathrm{C}[k/k_{fs}(t_{eq})]^2 $ where $C =\mathrm{O}(1)$ and independent of statistics for thermal
relics. We provide simple and accurate fits for $T(k)$ in a wide range
of \emph{small} scales $ k > k_{fs}(t_{eq})$ for thermal relics
  and different initial conditions.   The   numerical and analytic
results    for {\it arbitrary} distribution functions and  initial
conditions   allow an   assessment of  DM candidates through their
impact on structure formation.
\end{abstract}

\maketitle
\tableofcontents
\section{Introduction and Results}

The \emph{concordance} $\Lambda\mathrm{CDM}$ standard cosmological
model successfully explains   a wide range of highly precise
astrophysical and cosmological observations. Main ingredients
are an early stage of accelerated expansion   (inflation) a more recent stage of
accelerated expansion driven by dark energy,  and the
presence of dark matter (DM) composed of primordial particles which
are cold and collisionless at the time when the first structures
formed\cite{dodelson,kt}.

\medskip

In the cold dark matter (CDM)
model, structure formation proceeds in a hierarchical bottom up
approach: small scales become non-linear and collapse first and
their merger and accretion leads to structure on larger scales. This
is a consequence of the fact that CDM features negligible small
velocity dispersions leading to a power spectrum that favors small
scales. In this hierarchical scenario, dense clumps that survive the
merger process form satellite galaxies.

Large scale $\Lambda\mathrm{CDM}$ simulations seemingly lead to an overprediction of
satellite galaxies \cite{moore2} by almost an order of magnitude over
 the number of satellites that have been observed in Milky-Way
sized galaxies \cite{kauff,moore,moore2,klyp,will}. These
simulations   also yield a distinct prediction: virialized DM halos
should feature a  density profile that increases monotonically
towards the center \cite{frenk,dubi,moore2,bullock,cusps} such as
the Navarro-Frenk-White (NFW) profile \cite{frenk} or more general
  central density profiles $ \rho(r) \sim r^{-\beta} $ with
$ 1\leq \beta \lesssim 1.5 $ \cite{moore,frenk,cusps}. These profiles
accurately describe clusters of galaxies but indicate a divergent cusp at
the center of the halo.

There is, however, an accumulating body of observational
evidence \cite{dalcanton1,van,swat,gilmore,salucci,battaglia,cen,wojtak}
that seem to indicate that the density profile in the central
regions of dwarf galaxies is smooth leading to the suggestion that
the central regions feature smooth cores instead of cusps as
predicted by CDM.

\medskip

Warm dark matter (WDM) particles were invoked
\cite{mooreWDM,turokWDM,avila} as possible solutions to the above
mentioned discrepancies both in the overabundance of satellite
galaxies and as a mechanism to smooth out the cusped density
profiles predicted by CDM simulations into the  cored profiles
seemingly observed in dwarf-spheroidal satellite galaxies (dShps).
WDM particles feature a range of velocity dispersion in between the
CDM and hot dark matter (HDM) leading to free streaming scales that
smoothes out small scale features and could be consistent with core
radii of the dSphs.


Even if these problems of the CDM model find other astrophysical solutions, it
remains an important and intrinsically interesting question to assess the clustering properties of
 WDM  candidates, since these may emerge from extensions beyond the standard model of particle
 physics. Furthermore, it remains an important aspect to understand possible departures of the
 CDM paradigm at small scales provided by alternative particle physics candidates.


\medskip

The gravitational clustering properties of collisionless DM in the
linear regime are described by the   power spectrum of gravitational
perturbations and in particular the transfer function that converts
the primordial power spectrum into the power spectrum later in the
matter dominated era, which in turn,  is the input in the large
scale numerical simulations of galaxy formation. Free
streaming \cite{kt} of collisionless DM leads to a suppression of
the transfer function on length scales smaller than the free
streaming scale via Landau damping \cite{bondszalay}.

\medskip

Perturbations in a collisionless system of particles with
gravitational interactions is fundamentally different from fluid
perturbations in the presence of gravity. The (perfect) fluid
equations correspond to the limit of vanishing mean free path. In a
gravitating fluid, pressure gradients tend to restore hydrostatic
equilibrium with the speed of sound in the medium, and short
wavelength fluctuations are simple acoustic waves. For large
wavelengths, the propagation of pressure waves cannot halt
gravitational collapse on a dynamical time scale. The dividing line
is the Jeans length: perturbations with  shorter wavelengths
oscillate as sound waves, while perturbations with longer
wavelengths undergo gravitational collapse.

In a gas of collisionless particles with gravitational interaction
the situation is different since the mean free path is much larger
than the size of the system (Hubble radius) and the fluid
description is not valid. Instead, the collisionless
Boltzmann-Vlasov  equation for the distribution function must be
solved to obtain the dynamics of perturbations
\cite{peebles,bert,dodelson}.

Just as in the case of plasma physics, the linearized Boltzmann-Vlasov
equation describes \emph{collective} excitations \cite{plasmas}. In
the case of a collisionless gas with gravitational interactions
these collective excitations describe particles free-streaming in
and out of the gravitational potential wells of which they are the
 source. The damping of short wavelength \emph{collective}
 excitations is akin to \emph{Landau damping in
 plasmas} \cite{plasmas}: it is a result of \emph{dephasing} via phase
mixing when the particles are out of phase with the
 potential wells that they produce \cite{bt} and
 leads to the collisionless damping of the collective
 modes \cite{trema1}.

\medskip

  Gilbert \cite{gilbert} studied the linearized Boltzmann-Vlasov
 equation in a matter dominated cosmology for non-relativistic
 particles described by an (unperturbed) Maxwell-Boltzmann
 distribution function. This equation was cast as a Volterra integral equation
whose numerical solution revealed a
 limiting value of the wavevector above which density perturbations are
 Landau  damped and below which they grow via gravitational (Jeans)
 instability \cite{gilbert}.   The results were consistent with replacing the speed of sound
 by the Maxwellian velocity dispersion in the Jeans length [up to a
 normalization factor of $\mathcal{O}(1)$]. Gilbert's equations were
 solved numerically to study: (i) the collisionless damping of density fluctuations in an expanding
cosmology with massive neutrinos with an approximated Fermi-Dirac distribution function
\cite{bondszalay}; (ii) the dissipationless clustering of neutrinos
with a truncation of the Fermi-Dirac distribution
function and an  analytic fit to the numerical solution of the integral equation \cite{bran}.
These results were also used to analyze the linear regime \cite{turokWDM};
(iii) cosmological perturbations and massive light neutrinos \cite{bertwatts,bert},
and more recently  similar integral  equations were solved  approximately for thermal neutrino
relics \cite{ringwald}.

\medskip

DM may be composed of several species \cite{palazzo},
a possibility that can be   accommodated in most extensions of the Standard Model, with candidates
ranging from weakly interacting massive particles (WIMPs) \cite{kt} to axions \cite{kt} and sterile
neutrinos \cite{dw,este,kusenko}.
   DM candidates may be produced in the early Universe via different mechanisms and some of them are
   conjectured to decouple and freeze-out with non-thermal distribution functions. Sterile neutrinos
   are an example of this DM candidate \cite{dw,este,kusenko}. It is important to understand
the clustering properties of possible DM particles that decoupled \emph{in or out} of local thermal
equilibrium. Predictions for the effective free-streaming length and transfer function
are a necessary ingredient in the study of structure formation. The free-streaming length in
such \emph{mixture} was recently investigated in ref. \cite{freestream}
for \emph{arbitrary} distribution functions of relic particles
   that decoupled in or out local thermodynamic equilibrium (LTE).
   This study analyzed the Boltzmann-Vlasov equation in Minkowski space-time
    focusing on the marginally unstable collective modes. It
   revealed important (and intriguing) aspects of the free streaming lengths
   associated with the distribution function of the decoupled  particles.

\medskip

   A program that bridges the microphysics of the production,
   evolution and decoupling of DM particles with large scale
   structure formation begins by obtaining the distribution function
   of the DM particles from the dynamics of
   production and decoupling. The DM distribution function determines
   the abundance \cite{kt}, primordial phase space properties and generalized
   constraints on the DM candidate \cite{hogan,coldmatter} and their
   free streaming lengths \cite{freestream}. Results in this direction have been reported
\cite{coldmatter,kupesha}. The primordial phase space density of
particles  $ \mathcal D $  permits to obtain deep insights on DM
\cite{coldmatter}.  The phase space density can only decrease during
 gravitational dynamics via mergers and ``violent relaxation''\cite{tg}.
 Recent photometric and kinematic data on dwarf
spheroidal satellite galaxies in the Milky Way (dShps) and the
observed DM density today yield upper and lower bounds on the mass,
primordial phase space densities and velocity dispersion of the DM
candidates\cite{coldmatter}. Combining these constraints with recent
results from $N$-body simulations yield estimates on the mass of the
DM particles   in the {\bf range of a few keV} \cite{coldmatter}.

\medskip

The DM distribution function after freeze-out is the main ingredient to obtain the
   transfer function and the power spectrum in the linear theory. In
   principle the DM transfer function may be obtained by modifying
   the publicly available CMB anisotropy codes\cite{cmbfast,lewis} that treat baryons, photons and dark
   matter \cite{ma},  to account for the different distribution functions of
several species and range of parameters, masses and couplings.

 In practice, this is a computationally daunting
   problem: fairly complicated non-equilibrium
distribution functions for a variety of possible DM species
(axions, sterile neutrinos, etc) must be input in the codes
(with substantial modifications of the standard codes). The exploration
of the parameter space (masses, couplings and mixing angles)
   and their impact on the transfer function would require an
   enormous computational effort.

   After matter-radiation equality, the contribution from the
   baryon-photon fluid to the DM transfer function can only be a few
   percent\cite{dodelson}. DM only couples to the baryon-photon
   fluid via gravitational interaction and the most prominent
   feature of this coupling on the DM transfer function are baryon
   acoustic oscillations (BAO) on a scale corresponding to the sound
   horizon at recombination, corresponding today to about
   $150\,\mathrm{Mpc}$\cite{EH,HS,eisenstein}. Our goal is to
   understand the (DM) transfer function at much \emph{smaller}
   scales $\lambda \lesssim \,\mathrm{Mpc}$, at which the effect of the acoustic oscillations of the baryon-photon
   fluid (BAO) can be safely neglected.

\medskip

   {\bf Objectives:} The main goals of this article are the following:

    {\bf (a):}    to provide an
   \emph{analytic, accurate and simple} framework  to obtain the DM transfer function during
   matter domination for
general initial conditions and \emph{ arbitrary distribution
   functions} of relic DM particles that decoupled in or out of
   LTE and that are \emph{non-relativistic} after matter-radiation equality. Neglecting the
 gravitational coupling to the baryon-photon fluid entails that the DM transfer function will eventually require
 corrections  at the  few percent level. However,  we seek to gain understanding of robust features of $T(k)$ at
 small scales.

    {\bf(b):} to understand
   the   impact of the statistics of the relic particle on the small scale properties of the
   transfer function. More precisely, we study  which features of the distribution function
affect more prominently the power spectrum at small scales.

\medskip

\medskip

{\bf Summary of Results:}
   We study the linearized collisionless Boltzmann-Vlasov equation
   in the non-relativistic limit for particles that decoupled in or
   out of LTE with arbitrary (but isotropic) distribution
   functions  and general  initial conditions.

It is transformed into an integro-differential equation for density perturbations which
features non-local kernels that include \emph{memory of gravitational clustering} and
represent corrections to the fluid description.

The influence of this memory becomes more important at \emph{small scales}.
Fermi-Dirac (FD) and Bose-Einstein (BE) statistics result in \emph{long range} memory,
with the longer range for Bose-Einstein statistics. Distribution functions that favor small
momenta lead to longer range memory and result in an \emph{enhancement of the transfer function
at small scales}.

\medskip

We obtain an \emph{exact} expression   for $T(k)$ for \emph{arbitrary initial conditions and
distribution functions} and provide a   Fredholm expansion   that systematically includes
the effects of memory of gravitational clustering and higher moments of the distribution function,
which are corrections to the fluid approximation.

An exhaustive numerical study of thermal relics that decoupled relativistically, (Fermions (FD)
and Bosons, (BE)) or non-relativistically [WIMPs, (MB)] for different initial conditions reveal that
the  first two terms of the Fredholm series, given by eqn. (\ref{Tint2ndord}) provide an
accurate and simple approximation to the exact transfer function. These include explicitly the
corrections to the fluid description and the influence of statistics on small scales.
This approximation furnishes a useful tool to
extract the broad properties of the transfer functions for arbitrary
initial conditions and distribution functions.

\medskip

The approximate expression for the transfer function  given by
eq.(\ref{Tint2ndord})
features  three  main ingredients: {\bf (i):} the solutions of Jeans'
fluid equations, {\bf(ii:)} the free streaming solution of the
Boltzmann equation \emph{in absence of gravitational perturbations},
{\bf (iii:)} a non-local kernel that depends on the distribution
function of the decoupled particles, it defines the second
contribution in eq.(\ref{Tint2ndord}), includes \emph{memory}
of gravitational clustering and represents a correction beyond the fluid approximation.

The scale of suppression of  $T(k)$ is the
comoving free streaming wave-vector at matter-radiation equality, $k_{fs}(t_{eq})$,
where
\be \displaystyle
k_{fs}(t) = \sqrt{\frac{4 \, \pi \; \rho_{0M} \;
a(t)}{\left\langle \displaystyle
\frac{{\vec{p}}^{\, 2}}{m^2}
\right\rangle}} \;  \label{kfs1} ,
\ee
and the angular brackets refer to the average with the distribution
function of the decoupled particle, $ \rho_{0M} $ is the DM matter
density today and $ a(t) $ the scale factor.

 We find,
\bea
k_{fs}(t_{eq}) = \left\{
\begin{array}{l} \displaystyle{
   \frac{5.88}{ \mathrm{pc}} \; \Big( \frac{g_d}2 \;
\Big)^\frac13 \; \Big(\frac{m}{100\,\mathrm{GeV}}
\Big)^\frac12 \; \Big(\frac{T_d}{10 \,\mathrm{MeV}}  \;
\Big)^\frac12} ~~\mathrm{WIMPs} \\\displaystyle{
     0.00284\, \Big(
\frac{g_d}2\Big)^\frac13 \; \frac{m}{\mathrm{keV}} \;
[\mathrm{kpc}]^{-1} ~~\mathrm{FD~thermal~relics}} \\
  \displaystyle{   0.00317\, \Big(
\frac{g_d}2\Big)^\frac13 \; \frac{m}{\mathrm{keV}} \;
[\mathrm{kpc}]^{-1} ~~\mathrm{BE~thermal~relics}}
         \end{array} \right.
\eea
where $ g_d $ is the number of relativistic species at decoupling.

\medskip

For large scales  $ k \lesssim k_{fs}(t_{eq}) $,
  the contribution of the non-local memory kernel is
subleading, whereas for scales $ k > k_{fs}(t_{eq}) $  the memory of gravitational
clustering  and corrections beyond the fluid approximation    become important.

 FD and BE distribution functions lead to long range memory kernels with  Bose-Einstein statistics
 leading to the longest range. Longer range memory leads to an enhancement of the transfer function
 (and power spectrum) at \emph{small scales} as depicted in fig. \ref{fig:gamagrande}.
 For $ k > k_{fs}(t_{eq}) , \; (\gamma > \sqrt2) $
the difference in statistics of the decoupled particles becomes very
important

\medskip

At large scales $k \ll k_{fs}(t_{eq})$  the transfer function   $ T(k) $ in terms of the
variable $ \gamma = \sqrt2 \; k/k_{fs}(t_{eq}) $ has the  following
behaviour for all cases considered
\be \label{taluniI}
T(k) = 1 - \left( \frac{\gamma}{\gamma_0}\right)^2 + {\cal O}(\gamma^4)
\ee
where $ \gamma_0 $ is    the same for Fermi-Dirac, Bose-Einstein and Maxwell-Boltzmann distribution
 functions,  and  is given by eq.(\ref{gamacero}).

The transfer function for the three different cases (MB, FD and BE) can be compared   by
 parametrizing them in terms of
$ \gamma $. This comparison for different initial conditions is displayed in    fig. \ref{fig:gamagrande}.
Simple functional fits to $T(k)$ are provided on different regions of $k$ for various initial
conditions for thermal relics.

\section{  Gilbert's Equation}

After decoupling from the plasma, or freeze-out, the distribution
function of the decoupled particles obeys the collisionless
Boltzmann-equation. In absence of gravitational perturbations the
solution of this equation yields a distribution function of the
form \cite{kt,coldmatter}
\be \label{distfun}
 {f}_0(P_f(t),t) =  {f}_0\Bigg(P_f(t)\; \frac{a(t)}{a_d}\Bigg)
= {f}_0(p) \; ,
\ee
where $ P_f(t), \; a(t), \; a_d $ are the physical momentum, scale factor and
its value at decoupling respectively, and $ \vec{p} $ is the comoving
momentum. Consistently with isotropy we assume that $ f_0 $ is a
function of $ p=|\vec{p}| $.

During the matter dominated era, for  modes that are deep inside the Hubble radius
and DM particles that are non-relativistic the evolution of DM density perturbations
and gravitational perturbations is studied with the non-relativistic Boltzmann-Vlasov
equation\cite{peebles}. The non-relativistic case applies to all  comoving scales (physical scales today)
$\lambda \lesssim 170\,\mathrm{Mpc}$  which were inside the Hubble radius at matter-radiation
equality and DM particles with masses $m  > \mathrm{few}\,\mathrm{eV}$ that decoupled before
 matter-radiation equality.

 To linear order in perturbations
the distribution function of the decoupled particle and the
Newtonian gravitational potential are \cite{peebles,dodelson,bert}
\bea
f(\vec{p};\vec{x};t) & =
&  {f}_0(p)+ F_1(\vec{p};\vec{x};t) \label{pertdist} \\
\varphi(\vec{x},t)  & =
 & \varphi_0(\vec{x},t)+\varphi_1(\vec{x},t) \; , \label{pertgrav}
\eea
where $ {f}_0(p) $ is the unperturbed distribution function of the decoupled
particle,   $ \varphi_0(\vec{x},t) $ is the background gravitational
potential that determines the homogeneous and isotropic
Friedmann-Robertson-Walker metric and $ \vec{p},\vec{x} $ are comoving
variables.

The reader is referred to
refs. \cite{gilbert,bondszalay,bran,bertwatts,bert,dodelson} for
details on the linearization of the non-relativistic collisionless Boltzmann-Vlasov
equation.

   In conformal time $ \tau $ and
in terms of comoving variables $ \vec{p},\vec{x} $ the linearized
Boltzmann-Vlasov equation is \cite{gilbert,bondszalay,bran}
\be\label{boltzeq}
\frac1{a} \frac{\partial F_1}{\partial \tau} + \frac{\vec{p}}{m \;
a^2} \cdot \vec{\nabla}_{\vec{x}} F_1 - m \;
\vec{\nabla}_{\vec{x}}\varphi_1\cdot \vec{\nabla}_{\vec{p}} {f}_0 =0 \; ,
\ee
along with Poisson's equation,
\be\label{poissoneq}
\nabla^2_{\vec{x}} \varphi_1 = \frac{4 \, \pi \; G \; m}{a} \int
\frac{d^3p}{(2\pi)^3}  \; F_1(\vec{x},\tau) \; .
\ee
It is convenient \cite{gilbert,bran} to introduce a new time variable
$s$ related to the conformal time $ \tau $ by
\be \label{svar}
ds = \frac{d\tau}{a} \; ,
\ee
and to take spatial Fourier transforms of the
gravitational potential $ \varphi_1(\vec{x},\tau) $ and the
perturbation $ F_1(\vec{x},\tau) $ to obtain
\be \label{boltzeq2}
\frac{\partial F_1(\vec{k},\vec{p}\,;s)}{\partial s}
+ \frac{i \vec{k}\cdot
\vec{p}}{m} \; F_1(\vec{k},\vec{p}\,;s) -i \;  \vec{k}\cdot
\vec{\nabla}_{\vec{p}} {f}_0(p) \; a^2(s) \; \varphi_1(\vec{k},s) =0
\ee
where
\be \label{poisson2}
\varphi_1(\vec{k};s) = -\frac{4 \, \pi  \, G \; m}{k^2 \; a(s)}\int
\frac{d^3p}{(2\pi)^3}F_1(\vec{k},\vec{p};s)
\ee
Integrating from the initial time $ s = s_i $ to $ s $,
the Boltzmann-Vlasov equation (\ref{boltzeq2}) yields
\be\label{solboltzeq2}
F_1(\vec{k},\vec{p};s) = F_1(\vec{k},\vec{p};s_i) \; e^{-i \,
\frac{\vec{k}\cdot \vec{p}}{m} \; (s-s_i)} + i \; m  \; \vec{k}\cdot
\vec{\nabla}_{\vec{p}}  {f}_0(p) \int^s_{s_i}\, ds' \; e^{-i \,
\frac{\vec{k}\cdot \vec{p}}{m} \; (s'-s_i)} \; a^2(s')  \; \varphi_1(\vec{k},s') \; .
\ee
This equation can be turned into an integral
equation for the gravitational potential by multiplying both sides
by $ -4 \, \pi \, G  \; m/[k^2  \; a(s)] $,  integrating in $ \vec{p} $, and using the
relation (\ref{poisson2}). We obtain
\be \label{gil1}
\varphi_1(\vec{k};s) + i \; \frac{4 \, \pi \, G  \; m^2}{k^2  \; a(s)} \int \frac{d^3p}{(2\pi)^3}
\; \vec{k}\cdot \vec{\nabla}_{\vec p} f_0(p)  \; \int^s_{s_i} ds'\; e^{-i
\frac{\vec{k}\cdot \vec{p}}{m} \; (s-s_i)} \; a^2(s') \;
\varphi_1(\vec{k},s') = -\frac{4 \, \pi \, G  \; m}{k^2  \; a(s)} \int
\frac{d^3p}{(2\pi)^3}  \; F_1(\vec{k},\vec{p};s_i) \; e^{-i
\frac{\vec{k}\cdot \vec{p}}{m}  \; (s-s_i)} \; .
\ee
The right hand side of this equation, the inhomogeneity, is determined by the
first term in the solution eq.(\ref{solboltzeq2}) and describes the
\emph{free streaming} solution of the Boltzmann-equation in absence
of gravitational perturbations. This can be seen directly by
performing the inverse Fourier transform in $ \vec{k} $ to spatial
coordinates, which yields for the inhomogeneity the following form:
\be \label{Fgra}
\int \frac{d^3p}{(2\pi)^3} \; \mathcal{F}[\vec{x}-\frac{\vec{p}}{m}(s-s_i)]
\quad {\rm where} \quad
\mathcal{F}[\vec{x}] =  -\frac{4 \, \pi \, G  \; m}{a(s)} \int
 e^{i \; \vec{k}\cdot \vec{x}} \;  F_1(\vec{k},\vec{p};s_i) \; \frac{d^3k}{k^2} \; .
\ee
For a species that has become non-relativistic: $ \vec{p}/m=\vec{v} $ where
$ \vec{v} $ is the comoving velocity. Using the relation eq.(\ref{svar})
and $ d\tau = dt/a(t) $ we see that
\be \label{lfs}
\frac{p}{m} \; (s-s_i) = \int^t_{t_i} \frac{v(t')}{a(t')}  \; dt' \equiv l_{fs}(v;t)
\ee
where $ v(t) = v/a(t) $ is the physical
velocity of the non-relativistic particle and $ l_{fs}(v;t) $ is the
\emph{comoving} free streaming distance that the particle with
comoving velocity $ v $ has traveled during (comoving) time from $ t_i $
until $ t $ \cite{kt}. The physical distance is obtained by multiplying
the above expression by $ a(t) $. Therefore the inhomogeneity  eq.(\ref{Fgra})
in eq.(\ref{gil1}) is of the form
$$
\int \frac{v^2 \; dv}{2 \; \pi^2} \; \mathcal{F}[x-l_{fs}(v;t)] \; .
$$
The variable $ s $, related to conformal time
as in eq.(\ref{svar}), obeys the differential equation
\be\label{sofa}
\frac{ds}{da} = \frac1{a^3 \; H(a)} \; ,
\ee
where $ a $ is the scale factor and $ H $ the Hubble parameter. For matter
dominated cosmology
\be \label{Hofa}
H(a)= \frac{H_{0M}}{a^{\frac32}} \; ,
\ee
where
\be \label{hom}
H^2_{0M} =  \frac{8 \, \pi \; G}3 \; \rho_{0M} \equiv H^2_0 \;  \Omega_M  \; ,
\ee
with $ H_0 =   100 \; h \; \mathrm{km}\,\mathrm{sec}^{-1}\,
 \mathrm{Mpc}^{-1}$ is  the Hubble parameter \emph{today}
and $ \rho_{0M} $ is the matter density \emph{today} with
$ \Omega_M = 0.233 $ for DM.

Normalizing the scale factor to unity
\emph{today}, namely $ a(0)=1 $ and taking $ s_i=0 $ as initial value for the time variable $ s $,
eq.(\ref{sofa}) is integrated to yield
\be  \label{sofa2}
s = \frac2{H_{0M} \; a^{\frac12}_i}\left[1-\left(\frac{a_i}{a}\right)^{\frac12}\right] \; .
\ee
The initial value of the scale factor $ a_i $
 corresponds to the value at matter-radiation equality
$ a_i = a_{eq} = 1/(1+z_{eq}) $ with $ z_{eq} \simeq 3050 $.
It is convenient to introduce a dimensionless time variable $ u $ as
\be \label{sofu}
s =  \frac{2 \; u}{H_{0M} \; a_{eq}^{\frac12}} \; ,
\ee
so that
\be  \label{varu}
u= 1-\left(\frac{a_{eq}}{a}\right)^\frac12 = 1 - \sqrt{\frac{1+z}{1+z_{eq}}}
\quad , \quad
0\leq u \leq 1-a_{eq}^{\frac12} \simeq 0.982
\ee
and the scale factor in terms of the variable $ u $ is
 \be\label{aofu}
 a(u) = \frac{a_{eq}}{(1-u)^2} \; .
\ee

\subsection{The free streaming length.}

We note that during matter domination, the comoving free streaming
distance is given by
\be \label{adeplfs}
l_{fs}(t) = v \int^t_{t_i} \frac{dt'}{a^2(t')}  = \frac{v}{H_{0M}}\int^t_{t_i} \frac{da}{a^\frac32}=
 \frac{2 \; v}{H_{0M}} \left[ \frac1{\sqrt{a_{eq}}} - \frac1{\sqrt{a(t)}} \right] \; .
\ee
We see that the second term in $ l_{fs}(t) $ scales as $ 1/\sqrt{a(t)} $.
This observation will be important below to establish the redshift dependence of
the \emph{comoving} free streaming wave-vector.

Motivated by the expression for the free-streaming distance
eq.(\ref{lfs}) and (\ref{adeplfs}) and by the usual Jeans' wavevector for fluids
\cite{peebles}, we introduce the \emph{physical} free streaming
wavevector,
\be\label{kpfs}
k_{p,fs}(t) = \frac{k_{fs}(t)}{a(t)} \; .
\ee
The \emph{comoving} free streaming wavevector is defined as
\be\label{kfs}
k^2_{fs}(t)=\frac{3 \; H^2_{0M} \; a(t)}{2 \; \langle \vec{V}^2\rangle}
= a^2(t) \;  \frac{4 \; \pi  \; G  \; \rho_M(t)}{\langle \vec{V^2}(t)
\rangle} \quad , \quad \rho_M(t) = \frac{\rho_{0M}}{a^3(t)} \quad , \quad
\langle \vec{V^2}(t) \rangle = \frac{\langle \vec{V^2} \rangle}{a^2(t)} \; .
\ee
This definition of the free streaming wavevector is analogous to that of
the Jeans' wavevector in a fluid replacing the speed of sound
by $ \sqrt{\langle \vec{V^2}\rangle}$ (see the discussion below).

It is clear from this expression that the comoving
free streaming wave vector scales as
\be \label{kfsscal}
k_{fs}(t) = k_{fs}(0) \; \sqrt{a(t)}  \; ,
\ee
where, using eq.(\ref{kfs}),
\be \label{kfso}
k_{fs}(0) = \sqrt{\frac{3 \; H^2_{0M}}{2 \,\langle \vec{V}^2\rangle}} \; .
\ee
The comoving free streaming wavelength is given by:
\be \label{fss}
\lambda_{fs}(t) \equiv \frac{2 \, \pi}{k_{fs}(t)} \; .
\ee
There is a simple interpretation of
$ k_{fs}(t_i) $: consider the comoving free streaming distance that a
particle with comoving velocity $ v=\sqrt{\langle \vec{V}^2\rangle} $
travels from the initial time $ t_i =t_{eq} $ until today.
Setting $ s = 0 $ in eq.(\ref{adeplfs}) and neglecting the
second term since $ 1/\sqrt{a(0)} = 1 \ll 1/\sqrt{a_{eq}} \simeq 55.2 $, we find
\be \label{lfsv}
l_{fs}(v;0) = \frac{2 \, \sqrt{\langle \vec{V}^2\rangle}}{H_{0M} \;
a^\frac{1}{2}_i} \; .
\ee
Therefore,
\be \label{equiva}
k_{fs}(t_{eq}) = 0.0181 \; k_{fs}(0) = \frac{\sqrt6}{l_{fs}(v;0)} \; \Rightarrow \;
\lambda_{fs}(t_{eq}) =\sqrt{\frac23} \; \pi \; l_{fs}(v;0) \; ,
\ee
We note that the free streaming wavevector at $t_{eq}$ is related to the free streaming distance that
the particle has traveled between matter-radiation equality and today.

The matter density \emph{today} is
\footnote{We consider a species with a single internal degree
 of freedom, a different value $ g $ can easily be included in the final expressions.}
\be \label{rho0}
\rho_{0M} = m \; n_0 \quad , \quad n_0=\int \frac{d^3p}{(2\pi)^3} \;  {f}_0(p)  \; .
\ee
It also proves convenient to introduce the density perturbation
\be \label{h}
\Delta(\vk,s) = m \int \frac{d^3p}{(2\pi)^3} \; F_1(\vk,\vp\,;s) \; ,
\ee
related to the gravitational potential as
\be \label{fidelta}
\varphi_1(\vk;s)  \; a(s) = - \frac{4 \, \pi \; G}{k^2} \; \Delta(\vk;s) \; ,
\ee
and the gravitational potential normalized at the initial time
\be \label{tildefi}
\Phi(\vk,u) = \frac{\varphi_1(\vk,s)}{\varphi_1(\vk,0)} \; ,
\ee
where $ s(u) $ is given by eq.(\ref{sofu}).

The unperturbed distribution function $ \tilde{f}_0(p) $ is a dimensionless function,
and as such it can be written as a function of the ratio of the comoving
momentum $ p $ and the value of the decoupling temperature
\emph{today}  $ T_{d,0} $ and other dimensionless quantities,
such as the ratio of the mass of the particle to the
decoupling temperature, dimensionless coupling constants from the microphysics of decoupling
 etc., \cite{kt,coldmatter}, namely
\be
 f_0(p) \equiv  f_0(y;x_1,x_2\cdots)
\quad , \quad y = \frac{p}{T_{d,0}}\quad , \quad
x_1 =\frac{m}{T_d} \cdots \; ,
\ee
where  $ T_d $ is the decoupling temperature and $ T_{d,0} = T_d \; a_d $ is its value
\emph{today}. Using entropy conservation it follows that \cite{kt}
\be \label{T0d}
T_{d,0} = \Big(\frac2{g_d}\Big)^\frac13 \; T_{cmb} \; ,
\ee
where
$ T_{cmb}= 2.348 \times 10^{-4} $ eV is the CMB temperature
today and $ g_d $ is the effective number of relativistic
degrees of freedom at decoupling. For these distributions
$$
n_0 = T_{d,0}^3 \; \int_0^\infty \frac{dy}{2 \; \pi^2}  \;  y^2 \;  {f_0}(y) \; .
$$
Integrating by parts the momentum integral in the left hand side of eq.(\ref{gil1}),
 integrating over the angle $ \hat{k} \cdot \hat{p}=
 \cos\theta $ and  dividing both sides of the integral equation
 (\ref{gil1}) by the initial value
\be \label{phiini}
\varphi_1(\vk;0) = -\frac{4 \, \pi \; G  \; m}{k^2 \; a_{eq}} \int
\frac{d^3p}{(2 \, \pi)^3} \; F_1(\vec{k},\vec{p}\,;0) \; ,
\ee
the integral equation (\ref{gil1}) becomes
\be\label{gil3}
{\Phi}(k,u) - \frac6{\alpha} \; (1-u)^2  \; \int^u_0  \Pi[\alpha \; (u-u')] \;
\frac{{\Phi}(k,u')}{[1-u']^4} \; du' = (1-u)^2  \; I[\alpha \, u] \; .
\ee
where
\be \label{alfa}
\alpha \equiv \frac{2 \; k}{ H_0  \; \Omega^\frac12_M  \; a^\frac12_i } \;
\frac{T_{d,0}}{m} \; .
\ee
Using eq.(\ref{T0d}) and the value
\be\label{omedm}
\Omega_M \; h^2 = 0.105
\ee
for non-baryonic dark matter, $ \alpha $ becomes
\be\label{alfanum}
\alpha = 240 \; \Big(\frac2{g_d}\Big)^\frac13 \;
\frac{\mathrm{keV}}{m} ~~ k ~~ \mathrm{kpc} \; .
\ee
The cosmologically relevant range where this approach applies
goes from scales well inside the Hubble radius (where the non-relativistic
approximation is valid) till the smallest scales where the linearized
approximations are valid
\be\label{rangok}
(1000\,\mathrm{Mpc})^{-1} \lesssim k \lesssim (0.01\,\mathrm{Mpc})^{-1} \; .
\ee
Therefore, the range of the dimensionless variable $ \alpha $ results as
\be\label{rangalfa}
0.24 \; 10^{-3} \;  \Big(\frac2{g_d}\Big)^\frac13 \; \frac{\mathrm{keV}}{m}
\lesssim \alpha \lesssim 24 \;  \Big(\frac2{g_d}\Big)^\frac13 \; \frac{\mathrm{keV}}{m} \; .
\ee
As it is customary, we only consider distributions spherically symmetric
on $ \vk $. The non-local kernel in eqn. (\ref{gil3}) is given by
\be\label{sigma}
\Pi[z] = \int_0^\infty dy \; y \; \tilde{f}_0(y) \; \sin[yz]  \, ,
\ee where we introduced the normalized distribution function \be
\tilde{f}_0(y) \equiv \frac{ {f_0}(y)}{\int_0^\infty dy \;  y^2 \;  {f_0}(y)}~~;~~ \int_0^\infty y^2 \tilde{f}_0(y) dy =1  \; , \label{normaf0}\ee
and
\be \label{gorda}
I[\alpha \, u] =  \frac1{\int_0^\infty p^2  \; d p  \; F_1(k,p\,;0)}
\; \int_0^\infty p^2  \; d p  \; F_1(k,p\,;0) \;
\displaystyle{\frac{\sin\left(
\displaystyle{\frac{\alpha \; p \; u}{T_{d,0}}}
\right)}{\left(\displaystyle{\frac{\alpha \; p \; u}{T_{d,0}}}
\right)} }\,,
\ee
where we have assumed that $ F_1 $ does not depend on the direction
of $ \vec{p} $.

The inhomogeneity $ I[\alpha \, u] $ obeys the initial conditions
\be \label{iniI}
I[\alpha \, u=0] =1 \quad , \quad
\frac{d}{du}I[\alpha \, u]\Big|_{u=0}=0 \; .
\ee
The density perturbation normalized at the
initial time namely
\be  \label{tildedelta}
\delta(k\,,u) = \frac{\Delta(k\,;s)}{\Delta(k\,;0)}
\ee
is related to $ {\Phi}(k;\,u) $  [see eq.(\ref{fidelta}) and eq.(\ref{aofu})] by
\be  \label{fidelta2}
\Phi(k;\,u) = \frac{a_{eq}}{a(u)} \; \delta(k ,u) = (1-u)^2 \;
   \delta(k ,u) \; .
\ee
Then, from eq.(\ref{gil3})  $ \delta(k ,u) $ obeys the Volterra equation of the second kind
\be  \label{gil2}
\delta(k,u) - \frac6{\alpha}   \int^u_0 \Pi[\alpha \; (u-u')] \;
\frac{\delta(k,u')}{[1-u']^2} \; du' = I[\alpha \, u] \; .
\ee
which is   Gilbert's equation\cite{gilbert}.

The initial conditions of the
inhomogeneity eq.(\ref{iniI}) lead to the following initial conditions
on the gravitational potential and density perturbations
\bea
&& \delta(k,u=0) = 1 \quad , \quad
\frac{d}{du} \; \delta(k,u)\Big|_{u=0} = 0 \label{inidel}\\
&& {\Phi}(k,u=0) = 1 \quad , \quad
\frac{d}{du} {\Phi}(k,u)\Big|_{u=0} = -2 \; . \label{inifi}
\eea

\subsection{Exact results from   Gilbert's equation}

The long wavelength limit $ \alpha \rightarrow 0 $ of eq.(\ref{gil2}) affords an exact
solution. In this limit,
\be
{\displaystyle \lim_{\alpha \rightarrow 0}} \;
\frac1{\alpha} \; \Pi\big[\alpha(u-u')\big]= u-u'\quad , \quad
I[0]=1 \; ,
\ee
and $ (u-u') \; \Theta(u-u') $ is the Green's function of the differential
operator $ d^2/d^2u $. Therefore taking the second derivative with
respect to $u$ of the Volterra equation (\ref{gil2}), we obtain
\be\label{kcero}
\ddot{ {\delta}}(0,u) -\frac6{(1-u)^2} \; \delta(0,u)  =0 \; ,
\ee
the solution satisfying the initial conditions eq.(\ref{inidel}) is
\be\label{solkcero}
\delta(0,u) = \frac35 \; \frac1{(1-u)^2}+ \frac25 \; (1-u)^3 \; ,
\ee
which in terms of the scale factor eq.(\ref{aofu}) is given by
\be  \label{solofa}
\delta(0,u) = \frac35 \; \frac{a(u)}{a_{eq}} + \frac25 \;
\Big[\frac{a_{eq}}{a(u)}\Big]^\frac32 \; .
\ee
This is the usual long-wavelength solution for density perturbations in a
matter dominated cosmology. From the eq.(\ref{fidelta2}) the
long-wavelength limit of the gravitational potential is
\be\label{k0fi}
\Phi(0,u) = \frac35 + \frac25 \; (1-u)^5 \; .
\ee
We now show that $ \delta(k,u) $ behaves as $ 1/(1-u)^2 $ for all
values of $ \alpha $ when $ u\rightarrow 1 $.

Let us assume a general power law behavior for $ \delta(k,u) $ as $ u\rightarrow 1 $,
\be \label{asydel}
\delta(k,u) \stackrel{u\rightarrow 1}{=} A(\alpha) \;
(1-u)^{-\beta}\left[1+\mathcal{O}(1-u)\right] \; .
\ee
For $ u\rightarrow 1 $ the integral in eq.(\ref{gil2}) is dominated by the
region $ u' \sim u \sim 1$ , and we note that
\be\label{siguup}
\Pi[\alpha \; (u-u')]\stackrel{u'\rightarrow u}{=} \alpha \; (u-u') \; .
\ee
In this limit we obtain
\be \label{keruup}
\frac6{\alpha} \int^u_0 \frac{\Pi[\alpha \; (u-u')]}{[1-u']^2} \, \, (1-u')^{-\beta}\,du'
\stackrel{u \rightarrow 1}{=} \frac6{\beta(\beta+1)} \; (1-u)^{-\beta} \; .
\ee
Inserting this result in eq.(\ref{gil2}) we find two
solutions: $ \beta = 2,  -3 $ which are precisely the
singular and regular solutions found in the long-wavelength limit \cite{dodelson}.
Therefore, the dominant behavior for $ u \rightarrow 1 $ for
\emph{all} values of $ \alpha $ is
\bea
&&  \delta(k,u) \stackrel{u \rightarrow 1}{=}\frac{A(\alpha)}{(1-u)^2}
= A(\alpha) \; \frac{a(u)}{a_{eq}} \label{utoone}\\
&&  \Phi(k,u) \stackrel{u \rightarrow 1}{=} {A(\alpha)} \;
\label{fiutoone}
\eea
where in order to find the function $ A(\alpha) $ we have to integrate the Gilbert equation from
$ u = 0 $, where the initial conditions were defined, up to  $ u \to 1 $.
$ \Phi(k,u) $ is a \emph{slowly} varying function of $ u $ in the sense
that it is always finite as we see from eq.(\ref{fiutoone}).

\bigskip

{\bf The transfer function:} $ T(k) $ is defined from the limit \cite{dodelson}
\be \label{Tofk}
{\displaystyle \lim_{u\rightarrow 1}} \;
\frac{a_{eq}}{a(u)} \; \frac{\delta(k,u)}{\delta(k,0)} =
\lim_{u\rightarrow 1}\Phi(k,u) \; ,
\ee
where we have used the relation eq.(\ref{fidelta2}) and the initial condition
eq.(\ref{inidel}). It is customary to normalize the transfer function
so that $ T(k)\rightarrow 1 $ as $ k\rightarrow 0 $, obtaining
\be \label{Tofk2}
T(k) \equiv {\displaystyle \lim_{u\rightarrow 1}} \;
\frac{\Phi(k,u)}{\Phi(0,u)} = \frac53 \; {\displaystyle \lim_{u\rightarrow 1}} \;
(1-u)^2 \; \delta(k,u) = \frac53 \; \Phi(k,u=1) = \frac53 \; A(\alpha) \; .
\ee
A  numerical study of  $ \delta(k,u) $ and $ T(k) $   depicted in figs. \ref{fig:gilini}-\ref{fig:mbtofk2tini},
\ref{fig:thermalfermionIT}, \ref{fig:thermalbosonIT}-\ref{fig:gamagrande}   show that
   $ T(k) $ decreases with  $ k $ (or $ \alpha $). This is to be expected, $k=0$ corresponds to the mode that grows the fastest as
   it is the most gravitationally unstable, larger
   values of $k$ result in slower growth as a consequence of free streaming.
The behaviour of the $ k = 0 $ mode  is that   of cold dark matter eq.(\ref{solkcero}).

Therefore, $ T(k) $ is a measure  of the suppression of the $ k > 0 $ wave modes
and the $k$-scale of   suppression of $T(k)$ is necessarily related to the free-streaming wavenumber.

The final power spectrum $ P_f(k) $ is related to the initial one
$ P_i(k) $ as \cite{dodelson}
\be \label{powerspec}
P_f(k) = T^2(k) \; P_i(k) \; .
\ee
If perturbations do not grow or decay substantially during the prior,
radiation dominated phase, $ P_i(k) $ is nearly the inflationary
primordial power spectrum \cite{dodelson}.

\bigskip

  Gilbert's equation (\ref{gil2}) can be solved in a power series
in $ u $, appropriate for short times. We find from eq.(\ref{gorda})
\be\label{Iuchi}
I[\alpha \, u] \buildrel{u \to 0}\over=  1 - b \; \alpha^2 \; u^2 +{\cal O}(u^4)
\quad , \quad b \equiv \frac1{6 \; T_{d,0}^2} \;
\frac{\int_0^\infty p^4  \; d p  \; F_1(k,p\,;0)}{\int_0^\infty p^2  \; d p  \; F_1(k,p\,;0)}
\ee
Thus, we find from eqs.(\ref{Iuchi}) and (\ref{gil2}),
\be
\delta(k,u) \buildrel{u \to 0}\over= 1 + u^2 \;
\left[3 -  b \; \alpha^2  \right] + {\cal O}\left(u^4 \right)
\ee
Therefore $ \delta(k,u) $ starts out  by growing or decreasing for small
$ u $ depending on whether $ \alpha < \alpha_c $ or  $ \alpha > \alpha_c $,
respectively, where
$$
\alpha_c = \sqrt{\frac3{b}} \; .
$$
Namely, for $ \alpha < \alpha_c $ the mode $ \delta(k,u) $ is
gravitationally unstable from the start ($ u=0 $) while for  $ \alpha > \alpha_c $
the mode starts being gravitationally stable.
However,  due to the cosmological expansion {\bf } all modes redshift and eventually become
gravitationally unstable at some time $ u < 1 $.
As can be seen from eq.(\ref{utoone}), for $ u \to 1 $ {\bf all} modes are unstable.

At early times, $ \alpha_c $ determines the stability limit of
density perturbations.

We carried out an exhaustive numerical study for cold, hot and warm dark
matter (Maxwellian (MB) , BE and FD distributions).
We find for large $ \alpha $ that $ T(k) $ decreases exponentially
in the Maxwell-Boltzmann case while for both bosons and fermions
$ T(k) $ asymptotically decreases with a power law or exponentially
(see below secs. \ref{sfer} and \ref{sbos}).

At different scales, the suppression of $ T(k) $ is described by
a characteristic wave vector $ k_{char} $ which depends on the
initial conditions, the particle statistics and the regime of wavenumbers.

We obtain $ k_{char} $ from the numerical study for different cases and $ k $ regimes, in all cases
considered we find that   $  k_{char} \sim k_{fs}(t_{eq}) $.

\subsection{Thermal and Gilbert's initial conditions}

It remains to specify the initial perturbation $ F_1(k,p;u=0) $
of the distribution function. Although in principle this function should
be obtained from the full evolution through the prior radiation
dominated stage, in what follows we consider the physically
motivated case of adiabatic perturbations for which the perturbation
$ F_1(k,p;u=0) $ corresponds to a   temperature perturbation,
$$
T_{d,0} \rightarrow T_{d,0}\left[1+\frac{\Delta T(k)}{T_{d,0}}\right]
\; ,
$$
namely,
\be \label{adiapert}
F_1(k,p;u=0) =  T  \; \frac{df_0(p,T)}{dT} \; \frac{\Delta T(k)}{T} \; .
\ee
These are the initial conditions proposed in
refs.\cite{bondszalay,turokWDM}. Alternative initial conditions were
proposed by Gilbert \cite{gilbert}, who chose
\be \label{gilini}
F_1(k,p;u=0) = f_0(p)~C(k) \; ,
\ee
with some unspecified function $ C(k) $.

\subsection{Solution of the Gilbert equation in powers of
$ k $.}

We can solve the Gilbert equation (\ref{gil2}) in powers of $ \alpha^2 $,
[that is, powers of $ k^2 $, see eq.(\ref{alfa})]
\be\label{desal}
\delta(k,u) = \delta_0(u) + \alpha^2 \; \delta_1(u) + {\cal O}(\alpha^4) \; ,
\ee
where $ \delta_0(u) $ is given by eq.(\ref{solkcero}) and $ \delta_1(0) = \delta'_1(0) = 0 $.
Inserting eq.(\ref{desal}) in eq.(\ref{gil2}), using eq.(\ref{Iuchi}) and deriving twice
with respect to $ u $ leads to the equation
$$
\left[\frac{d^2}{du^2} -\frac6{(1-u)^2} \right]\delta_1(u) = 2 \; \left[3 \, b - {\bar b} -
3 \; b \; \delta_0(u)\right]
$$
where $ {\bar b} = b $ for Gilbert initial conditions and $ {\bar b} = \frac53 \; b $
for temperature initial conditions. This inhomogeneous  differential equation for
$ \delta_1(u) $ can be easily solved using the Green's function of the differential
operator in the l. h. s. We find for $ u \to 1 $ after calculation,
\bea
\delta_1(u) \buildrel{u \to 1}\over= \left\{
\begin{array}{l} \displaystyle{
- \frac{8 \, b}{35} \; \frac1{(1-u)^2} + {\cal O}(1)} \quad {\rm Gilbert ~ initial ~ conditions,}
\\ \\ \displaystyle{
- \frac{31 \, b}{105} \; \frac1{(1-u)^2} + {\cal O}(1)}\quad {\rm Temperature ~ initial ~ conditions.}
\end{array} \right.
\eea
Using eq.(\ref{Tofk2}) we now obtain $ T(k) $ in powers of $ \alpha^2 $,
\bea \label{Talf2}
T(k) = \left\{
\begin{array}{l} \displaystyle{
1- \frac{8 \, b}{21} \; \alpha^2 + {\cal O}(\alpha^4)} \quad {\rm Gilbert ~ initial ~ conditions,}
\\ \\ \displaystyle{
1- \frac{31 \, b}{63} \; \alpha^2 + {\cal O}(\alpha^4)}\quad {\rm Temperature ~ initial ~ conditions,}
\end{array}\right.   
\eea where $b$ is given by eqn. (\ref{Iuchi}).

\subsection{  Gilbert's equation as an integro-differential equation.}

The results obtained above suggest to convert the Volterra equation
into an integro-differential equation.  Taking the second derivative
with respect to $ u $ of   eq.(\ref{gil2}) yields
\be
\ddot{\delta}(k,u) - \frac{6 \; \delta(k,u)}{(1-u)^2}+6 \; \alpha^2
\int^u_0 du' \int_0^\infty  dy \;  y^4  \;
\tilde{f}_0(y)   \; \frac{\sin[\alpha \;
y \; (u-u')]}{\alpha \; y} \; \frac{\delta(k,u')}{(1-u')^2} = \ddot{I}[k,u] \; .
\ee
It proves convenient to write the following identity in the $ y $-integral:
\be \label{repla}
y^4 = y^2  \; \overline{y^2} + y^2 \; (y^2-\overline{y^2} ) \; ,
\ee
where
\be \label{overy}
\overline{y^2} = \int_0^\infty  dy  \; y^4  \; \tilde{f}_0(y)
\quad {\rm with} \quad \int_0^\infty  dy  \; y^2  \; \tilde{f}_0(y) = 1 \; ,
\ee
and to introduce
\be \label{gamma}
3 \; \gamma^2 \equiv \alpha^2  \; \overline{y^2}  \quad ,  \quad
\gamma =  \frac{2 \, k \; T_{d,0}}{m \; H_0} \;
\sqrt{\frac{\overline{y^2}}{3 \; \Omega_M \; a_{eq}}} \; ,
\ee
where eq.(\ref{alfa}) was used.

In terms of $ k_{fs} $ [see eqs.(\ref{hom}), (\ref{kfso}) and (\ref{gamma})]  we find that
\be \label{gammarat}
\gamma^2 = \frac{2 \, k^2}{k^2_{fs}(t_{eq})} = \frac{2\,k^2}{k^2_{fs}(0) \;a_{eq}}
\quad ,  \quad
k_{fs}(0) = \sqrt{\frac{3 \; \Omega_M}{2 \; \overline{y^2}}} \; H_0 \;
\frac{m}{T_{d,0}} \; .
\ee
Using the original eq.(\ref{gil2}) to replace the term with
$\overline{y^2}$ in terms of $ \td(k,u) $ and $ I[\alpha \, u] $, we obtain
the following integro-differential equation for density perturbations
\be \label{diffeqn}
\ddot{\td}(k,u) - \frac{6 \; \td(k,u)}{(1-u)^2}
+ 3 \; \gamma^2  \; \td(k,u) - \int^u_0 du' \; K(u-u') \;
\frac{\td(k,u')}{(1-u')^2} = \ddot{I}[k,u] + 3 \; \gamma^2  \; I[\alpha \, u] \; ,
\ee
where the non-local kernel $ K(u-u') $ is given by
\be\label{kernel}
K(u-u') =  6 \, \alpha \int_0^\infty y \;
(\overline{y^2}-y^2) \; \tilde{f}_0(y) \;  \sin[\alpha \; y \; (u-u')] \; dy  \; .
\ee
and we note that
\be \label{V2}
\overline{y^2} \; \Big(\frac{T_{d,0}}{m}\Big)^2
= \langle \vec{V}^2 \rangle
\ee
is the three dimensional velocity dispersion  of the
non-relativistic particles \emph{today}.

\medskip

Using eq.(\ref{lfsv}), $ \gamma $ [eq.(\ref{gamma})] can also be written as
\be  \label{otrogama}
\gamma = \frac{\sqrt2 \; k}{k_{fs}(t_{eq})} =
\frac1{\sqrt3} \; k \; l_{fs}(v;0) =
  \frac{2 \, \pi}{\sqrt3} \; \frac{l_{fs}(v;0)}{\lambda} \; ,
\ee
where $ \lambda $ is the wavelength of the perturbation.

Using eqs.(\ref{kfso}), (\ref{T0d}) and (\ref{omedm}),
$ k_{fs}(t_{eq}) $ and $ l_{fs}(v;0) $ are given by
\bea \label{kofs}
k_{fs}(t_{eq})  & = & \frac{0.0102}{\sqrt{\overline{y^2}}} \;
\Big(\frac{g_d}2\Big)^\frac13 \;
\frac{m}{\mathrm{keV}} \; [\mathrm{kpc}]^{-1} \\
\lambda_{fs}(t_{eq}) = \sqrt{\frac23} \; \pi \;
l_{fs}(v;0)  & = &  616 \; \sqrt{\overline{y^2}} \;
\Big(\frac2{g_d}\Big)^\frac13 \; \frac{\mathrm{keV}}{m} ~~
\mathrm{kpc} \label{lfsn} \; .
\eea
The transfer function for the three different cases (MB, FD and BE)
turns to be parametrized by the dimensionless ratio
$ \gamma = \sqrt2 \; k/k_{fs}(t_{eq}) $ allowing us to compare the three different cases,
in fig. \ref{fig:gamagrande} below.

The alternative form eq.(\ref{diffeqn}) of the original Volterra
(Boltzmann-Vlasov) [eq.(\ref{gil2})] equation has several merits :
\begin{itemize}

\item{Neglecting the non-local term and the inhomogeneity,
eq.(\ref{diffeqn})
can be written in a more familiar form in terms of cosmic time $t$:
\be\label{fluid1}
\frac{d^2 \td}{dt^2} + 2 \, H \; \frac{d\td}{dt} +\left[\frac{k^2 \;
\langle \vec{V}^2\rangle}{a^4(t)}-4 \, \pi \; G \; \rho_M(t)\right]\td =0
\ee
where $ \langle \vec{V}^2\rangle $ is given by eq.(\ref{V2})
and $ \langle \vec{V}^2\rangle/a^2(t) $ is the
\emph{physical} squared velocity of the non-relativistic particles.

The term between brackets in eq.(\ref{fluid1}) can be written as
\be
\frac{\langle \vec{V}^2\rangle}{a^2(t)} \Bigg[\frac{k^2}{a^2(t)} -
\frac{k^2_{fs}(t)}{a^2(t)} \Bigg] \label{kjean}
\ee
where $ k_{fs}(t) $ is the \emph{comoving} free streaming wavevector
eq.(\ref{kfs}). This equation must be compared to the usual fluid
equation \cite{peebles} in which the comoving Jeans wave vector
$ k_J(t) $ replaces $ k_{fs}(t) $ with
\be \label{kJdef}
k^2_J(t)= \frac{4\pi \;  G}{c^2_s} \;  \rho_{0M} \; a(t) \; ,
 \ee
where $ c_s $ is the comoving speed of sound during matter domination.
This expression shows that the \emph{comoving} Jean's wavevector in the
fluid description scales as $ \sqrt{a(t)} $ during matter
domination, just as the free streaming wavevector $ k_{fs}(t) $ does.
Thus, we see that the long-wavelength limit $ \alpha \rightarrow 0 $ ($ \gamma
\rightarrow 0 $) of eq.(\ref{diffeqn})  yields
the familiar fluid description.

Therefore the contribution from the non-local kernel $K(u-u')$ is a \emph{correction} to the
fluid description. It will be seen below that this correction becomes important at small scales $k > k_{fs}(t_{eq})$.

\item{In the long wavelength limit $ \alpha \rightarrow 0 $ ($ \gamma
\rightarrow 0 $) the kernel $ K(u-u') $ decreases as
$ \gamma^4 \ll \gamma^2 $ and its contribution to the dynamics of density
perturbations becomes negligible.}

Furthermore, expanding the kernel $ K(u-u') $ in
a power series in $k$ it can be seen that each term corresponds to
higher correlations of the form
$$
\left\langle \left(\frac{p}{m}\right)^{2+n} \right\rangle - \left\langle
\left(\frac{p}{m}\right)^{n} \right\rangle \; \left\langle \left(\frac{p}{m}\right)^2
\right\rangle \; .
$$
These correlations are typically neglected in the
moment-hierarchy of the Boltzmann equation \cite{dodelson}.}
\item{At short times $ u \sim 0 $ it follows that
$$
\int^u_0 K(u-u') \;  du' \buildrel{u \to 0}\over= {\cal O}
\left(\gamma^4 \; u^4 \right) \; ,
$$
hence, its contribution is subleading for $ u \lesssim 1/\gamma $
since   $ \td(k,u\sim 0) \sim 1 $ by the initial condition.}
\item{As $ u\rightarrow 1 $, the $ u' $ integral in the kernel is
dominated by the region $ u\sim u' \sim 1 $ for which $ \td(k,u')
\sim  A(\alpha)/(1-u')^2 $ as obtained in eq.(\ref{utoone}),
therefore it follows that for $ u\sim 1 $,
\be
\int^u_0 du' \; K(u-u')  \; \frac{\delta(k,u')}{(1-u')^2}
\buildrel{u \to 1}\over=\alpha^4 \; [\overline{y^4}-(\overline{y^2})^2]
\ln\frac1{1-u} \ll \frac1{(1-u)^2} \; .
\ee
Therefore in this region the
contribution of the kernel is subleading as compared to the first
three terms in eq.(\ref{diffeqn}) although it could be larger than
the inhomogeneity on the right hand side.}
\item{It explicitly yields the asymptotic behavior
$ \td(k,u\rightarrow 1) \propto 1/(1-u)^2 $ as a consequence of the
second term in eq.(\ref{diffeqn}) and the analysis above. This feature
will be seen manifestly in the analysis below.}
\end{itemize}

\subsection{The \emph{exact} transfer function and its Fredholm expansion.}

The above analysis points out that in all the regimes of interest
the non-local contribution from the kernel is small as compared to the
first three terms in the eq.(\ref{diffeqn}) and can be treated
perturbatively. For this purpose, it is convenient to write
eq.(\ref{diffeqn}) as
\be \label{differ}
\ddot{\delta}(k,u)-\frac6{(1-u)^2} \; \delta(k,u)+
3 \; \gamma^2 \; \delta(k,u) = S[\delta;u] \; .
\ee
The source term is
\be\label{source}
S[\delta;u] = S_B[u]+S_{NB}[\delta;u]
\ee
where $ S_B[u] $ and $ S_{NB}[\delta;u] $ play the role of the
Born and next to Born approximations,
\bea \label{S0}
S_B[u] & = & \ddot{I}+3 \, \gamma^2 \; I  \\ \label{S1}
S_{NB}[\delta;u] & =& \int_0^u du' \; K(u-u') \;
\frac{\delta(k,u')}{[1-u']^2} \; .
\eea
As discussed above [see eq.(\ref{fluid1})], in cosmic time
the left hand side of eq.(\ref{differ}) is
precisely of the form of the Jeans equation for fluids. The right hand
side $ S[\delta;u] $
may be interpreted as an additional pressure term non-local
in time which includes free-streaming through the initial condition
in $ S_B $. In the long wavelength limit $ S[\delta;u] $ vanishes
along with the usual pressure term $ 3 \, \gamma^2 \; \delta(k,u) $
leading to the usual fluid-like equation for cold-dark matter density
perturbations.

Eq.(\ref{differ}) lends itself to a Fredholm (iterative)
solution, the basis of which is the homogeneous solution and the
retarded Green's function of the differential operator on the left
in eq.(\ref{differ}).

For $ S[\td,u]=0 $, eq.(\ref{differ}) becomes the homogeneous
differential equation
\be \label{homdif}
\ddot{\delta}^{(0)}(k,u)-\frac6{(1-u)^2} \; \delta^{(0)}(k,u)
+ 3 \; \gamma^2 \; \delta^{(0)}(k,u) = 0 \; .
\ee
which is solved in terms of Bessel's functions. The general solution is given by
\be \label{genh}
\delta^{(0)}(z)= A~h_1(z)+ B~h_2(z) \; ,
\ee
with
\be
z \equiv z_0 \; (1-u) \quad , \quad
z_0 = \sqrt3 \; \gamma \; , \label{zetavar}
\ee
$ h_1(z) $ and $ h_2(z) $ are related to the spherical
Bessel functions $ n_2(z) $ and $ j_2(z) $ \cite{MF},
\bea
&&h_1(z)\equiv - z \; n_2(z)  =
\left(\frac{3}{z^2}-1\right)\cos z + \frac{3}{z}
\sin z \; ,\label{h1}\\
&& h_2(z) \equiv  z \; j_2(z) =
\left(\frac{3}{z^2}-1\right)\sin z - \frac{3}{z} \cos z\; ,
\label{h2}
\eea
$ h_1(z) $ and $ h_2(z) $ are the fundamental irregular and regular solutions, respectively,
with the small argument
behavior for $ z\rightarrow 0 $ ($ u \rightarrow 1 $):
\be\label{h1asy}
h_1(z) \buildrel{z \to 0}\over= \frac3{z^2}  \quad , \quad
h_2(z) \buildrel{z \to 0}\over= \frac{z^3}{15} \; ,
\ee
and  Wronskian
\be\label{w}
W[z] = h'_2(z) \; h_1(z)-h'_1(z) \; h_2(z) =1 \; ,
\ee
where the prime denotes $ d/dz $. The mode functions $ h_{1,2}(u) $ are
identified as the growing and decaying solutions of Jeans's fluid
equations (\ref{fluid1}). The initial conditions eq.(\ref{inidel}) yield
\be\label{hini}
\td^{(0)}(z_0)  = 1 \quad , \quad \td^{(0)'}(z_0) =0 \; .
\ee
Using eq.(\ref{w}) we find
\be \label{coefs}
A = h'_2(z_0) \quad , \quad B= -h'_1(z_0) \; ,
\ee
and the homogeneous solution is given by
\be\label{homo}
\td^{(0)}(z) = h'_2(z_0) \; h_1(z)-h'_1(z_0) \; h_2(z) \; .
\ee
Using the small argument behavior eq.(\ref{h1asy}), for $ \gamma=0 $
($ k=0 $) the homogeneous solution $ \td^{(0)}(z) $ reduces to eq.(\ref{solkcero}).

\medskip

{\bf The general solution:} The inhomogeneous equation
(\ref{differ}) can be formally solved in terms of the retarded
Green's function of the differential operator on its left hand side
which obeys
\be\label{ecdig}
\left[\frac{d^2}{du^2} -\frac6{(1-u)^2} + 3 \; \gamma^2\right]
G(u,u') = \delta(u-u') \; .
\ee
$ G(u,u') $ can be explicitly written in terms of  $ h_{1,2}(u) $ [eqs.
(\ref{h1})-(\ref{h2})] as,
\be \label{green}
G(u,u') = \frac1{\sqrt3 \; \gamma} \;
\left[h_1(u) \; h_2(u')-h_2(u) \; h_1(u')\right] \; \Theta(u-u') \; .
\ee
The formal solution of eq.(\ref{differ}) is then given by
\be \label{integeqn}
\td(k,u)= \td^{(0)}(z)+\frac1{\sqrt3 \; \gamma}\int_0^u
\left[h_1(u) \; h_2(u')-h_2(u) \; h_1(u')\right] \; S[\td;u'] \; du'
 \; .
\ee
It is convenient to separate explicitly the
source term $ S_B[u] $ eq.(\ref{S0}) which does not involve the kernel
$ K(u-u') $, but only involves the free streaming solution of the
Boltzmann equation in absence of self-gravity. Integrating by parts twice
the term with $ \ddot{I} $, using the differential equation (\ref{ecdig})
obeyed by the Green's function and the initial conditions
eq.(\ref{iniI}), we obtain
\be \label{inteq}
\td(k,u)= \td^{(1)}(k,u)+\frac1{\sqrt3 \; \gamma}\int_0^u
\left[h_1(u) \; h_2(u')-h_2(u) \; h_1(u')\right] \; S_{NB}[\td;u']
\; du' \; ,
\ee
where
\be \label{td1}
\td^{(1)}(k,u) = I[\alpha \, u] + \frac6{\sqrt3 \; \gamma} \int_0^u
\left[h_1(u) \; h_2(u')-h_2(u) \; h_1(u')\right] \;
\frac{I[\alpha \, u']}{(1-u')^2} \;   du' \; .
\ee
The gravitational potential $ \Phi(k,u) $ can be analogously expressed
using its relation with the density perturbation $ \delta $
eq.(\ref{fidelta2}) and
eq.(\ref{integeqn}) with the result
\be \label{inteqfi}
\Phi(k,u)= \Phi^{(1)}(k,u)+\frac1{\sqrt3 \; \gamma} \; (1-u)^2 \; \int_0^u
\left[h_1(u) \; h_2(u')-h_2(u) \; h_1(u')\right] \;
S_{NB}[\td;u'] \; du'  \; ,
\ee
where
\be \label{tfi1}
\Phi^{(1)}(k,u) = (1-u)^2  \; \td^{(1)}(k,u)  \; ,
\ee
is the Born term.

It is straightforward to check that for $ k=0 $ the second
term in eq.(\ref{inteqfi}) vanishes and $ \Phi^{(1)}(\vec{0},u) $ is
given by eq.(\ref{k0fi}) (for this note that $ I[0]=1 $).

Free streaming for $ k \neq 0 $ leads to collisionless Landau damping of
the gravitational potential and density perturbation. Although
density perturbations still grow as $ 1/(1-u)^2 $ for $ u\rightarrow 1 $
the gravitational potential is bound $ |\Phi| \leq 1 $ and is a
slow variable.

We obtain an \emph{exact} expression for $ T(k) $ inserting eqs.(\ref{inteqfi})
into the definition of the normalized transfer function eq.(\ref{Tofk})
and using the small argument limits eq.(\ref{h1asy}),
\be \label{Tint}
T(k)= \frac{10}{\sqrt3 \; \gamma^3 } \int^1_0 h_2(u) \;
\Bigg[\frac{I[\alpha \, u]}{(1-u)^2}+\frac16 \; S_{NB}[\td;u] \Bigg] \; du \; ,
\ee
where $ \td(k,u) $, argument of $ S_{NB} $ [see eq.(\ref{S1})], is the solution of the integral equation
(\ref{inteq}) and we have neglected terms proportional to $ a_{eq} \sim 10^{-4} $.
In obtaining this result we have used the fact that $ I[\alpha \, u] $
given by eq.(\ref{gorda}) is finite at $ u = 1 $ and that in this limit
$ \td(k,u) \propto 1/(1-u)^2 $. Only the terms featuring $ h_1(u) $
outside the integrals survive in this limit. The terms with $ h_1(u') $
inside the integrals yield at most terms $ \propto 1/(1-u)^3 $ but they
are multiplied by $ (1-u)^5 $, and vanish for $ u \rightarrow 1 $.
Furthermore, in the long wavelength limit $ I[0] =1~;~
S_{NB} \propto \gamma^4 $ and from the small argument behavior eq.(\ref{h1asy}) it is
straightforward to confirm that $ T(k=0)=1 $.

\bigskip

The integral equations (\ref{inteq}) and (\ref{inteqfi}) can be iterated generating
the usual Fredholm series as
\bea \label{FS}
&& \td(k,u)= \td^{(1)}(k,u)+ \td^{(2)}(k,u)+\cdots \; , \cr \cr
&&\Phi(k,u) = \Phi^{(1)}(k,u)+\Phi^{(2)}(k,u) +\cdots \label{fifred} \; ,
\eea
where $ \td^{(1)}(k,u) $ is given by eq.(\ref{td1}) to which we refer as
the \emph{Born} term because of its obvious similarity to the solution
of a potential scattering problem,
\bea \label{td2}
\td^{(2)}(k,u) & = & \frac{1}{\sqrt3 \; \gamma}\int_0^u
\left[h_1(u) \; h_2(u')-h_2(u) \; h_1(u')\right] \; S_{NB}[\td^{(1)};u']
\; du'\\ \label{S11}
S_{NB}[\td^{(1)};u] & = &  \int_0^u
du' \; K(u-u') \; \frac{\delta^{(1)}(k,u')}{[1-u']^2} \; ,
\eea
$ \Phi^{(1)}(k,u) $  is given by eq.(\ref{tfi1}) and
\be \label{fifred2}
\Phi^{(2)}(k,u) =  (1-u)^2 \;  \frac1{\sqrt3 \; \gamma} \int_0^u
\left[h_1(u) \; h_2(u')-h_2(u) \; h_1(u')\right] \; S_{NB}[\td^{(1)};u'] \; du'
 \; .
\ee
The small argument behavior eq.(\ref{h1asy}) makes manifest that the
solution $ \td(k,u) $ indeed has the
asymptotic behavior $ \propto 1/(1-u)^2 $ as $ u\rightarrow 1 $.
Furthermore, since the function $ I[\alpha \, u] $ describes the
free-streaming solution of the Boltzmann equation in absence of
self-gravity, the formal solution eq.(\ref{inteq}) exhibits explicitly
the free-streaming phenomenon.

\medskip

The transfer function has also a systematic Fredholm expansion which up to second order yields
\be \label{Tint2ndord}
T(k)= \frac{10}{\sqrt3 \; \gamma^3 }
\int^1_0 h_2(u) \; du  \; \Bigg[\frac{I[\alpha \, u]}{(1-u)^2} + \frac16 \;
S_{NB}[\td^{(1)};u] \Bigg] \equiv T_B(k) + T_{NB}(k) \; ,
\ee
where we refer to the first order term
\be \label{Tborn}
T_B(k) = \frac{10}{\sqrt3 \; \gamma^3}
\int^1_0 h_2(u) \;  du  \; \frac{I[\alpha \, u]}{(1-u)^2} \; ,
\ee
as the \emph{Born} term because its origin is the Born term for the
gravitational potential $ \Phi^{(1)}(k,u) $ and
\be \label{T2ndord}
T_{NB}(k)= \frac5{3 \, \sqrt3 \; \gamma^3} \int^1_0 h_2(u) \;  S_{NB}[\td^{(1)};u]
 \; du \; ,
\ee
is the next to Born (second order) correction, where $ \delta^{(1)}(k;u) $
and $ S_{NB}[\delta^{(1)};u] $ are given by eqs.(\ref{td1}) and
(\ref{S11}), respectively.

\medskip

At this point it is worthwhile to emphasize the following important
aspect: the first order, Born term in the transfer function only
depends on the initial condition and describes the free-streaming
suppression. The second order term, the
integral of $ S_{NB} $, contains the information on the higher moments of
the \emph{distribution function} of the decoupled particles and the corrections to the
fluid approximation through
the non-local kernel $ K(u-u') $ eq.(\ref{kernel}). The
details of the distribution function beyond the first moment
$ \langle\vec{p}^{\,2}\rangle $ enter the transfer function  at second
order and beyond in the Fredholm series. It will be shown below
that these contributions  include  \emph{memory of gravitational
clustering} and become important at short wavelengths $\gamma > 1$.

\medskip

Eq.(\ref{Tint2ndord}) provides a useful approximation
to the transfer function for general initial conditions and
distribution functions.
Specific examples for which we obtain an exact numerical evaluation
(see below) show that the second order approximation
eq.(\ref{Tint2ndord}) is {\it remarkably accurate} in a
wide range of scales. These are some of the main results of this article.

\section{Cold Dark Matter: Non-relativistic, Maxwell-Boltzmann
distribution. }

The unperturbed distribution function for particles that decoupled
while non-relativistic in local thermal equilibrium is a solution of
the collisionless Boltzmann-equation (in absence of gravitational
perturbations), given by
\be \label{decNR}
f_0(P_f(t),t) = \mathcal{N} \;
e^{-\frac{P^2_f(t)a^2(t)}{2\,m\,T_d a^2_d}} \; ,
\ee
where the explicit expression for the normalization factor $ \mathcal{N} $
may be found in ref.\cite{coldmatter},  $ P_f(t) $ is the physical
momentum, $ a(t) $ is the scale factor and $ T_d, \; a_d $ are the
temperature and scale factor at decoupling respectively.
Since $ p=P_f(t) \; a(t) $ is
the comoving momentum (equal to the physical momentum \emph{today})
and $ T_d  \; a_d = T_{d,0} $ is the decoupling temperature \emph{today},
the unperturbed Maxwell-Boltzmann distribution function
can be written in terms of dimensionless variables as
\be\label{MBdist}
f_0(y;x) = \mathcal{N} \; e^{-\frac{y^2}{2 \, x}} \quad ,  \quad
y = \frac{p}{T_{d,0}}\quad ,  \quad x = \frac{m}{T_d}\,,
\ee and the normalized distribution function is given by
\be\label{tildefMB}
\tilde{f}_0(y) =  \frac4{\sqrt{\pi} \; [2 \, x]^\frac{3}{2}} \;
e^{-\frac{y^2}{2 \, x}} \; ,
\ee
$ \overline{y^2} $ [eq.(\ref{overy})] and $ \alpha $ [eq.(\ref{gamma})]
are in this case,
\be
\overline{y^2} =  3 \; \frac{m}{T_d}  \quad , \quad
\alpha = \sqrt{\frac{T_d}{m}} \; \gamma \; . \label{overy2MB}
\ee
The typical values of the masses and decoupling temperatures for WIMPs
are $ m \sim 100\,\mathrm{GeV} $ and $ T_d \sim 10 $ MeV
\cite{dominik}, for which $ g_d \sim 10 $ \cite{kt}.
Using the result for the free-streaming wavevector today eq.(\ref{kofs})
we find
\be
k_{fs}(t_{eq}) =
\frac{5.88}{\mathrm{pc}} \; \Big( \frac{g_d}2
\Big)^\frac13 \; \Big(\frac{m}{100 \,\mathrm{GeV}}
\Big)^\frac12 \; \Big(\frac{T_d}{10 \,\mathrm{MeV}}
\Big)^\frac12  \; .\label{kofsCDM}
\ee
Within the cosmologically relevant range of scales where the linearized approximation is valid,
eq.(\ref{rangok}) and  using (\ref{gamma}) it follows  that in this range  $ \gamma \leq 10^{-5} $.

\medskip

The kernel $ \Pi[\alpha \; (u-u')] $ that enters in the Volterra equation
(\ref{gil2}) is best obtained by performing the momentum integrals
of the distribution function in cartesian coordinates rather than
performing the angular integration first. This is a consequence of
the fact that the distribution function is a function of
$ \vec{p}^{~2} $. We obtain  the following   equation for
density perturbations
\be \label{gileq}
\td(k,u)-6 \;  \int_0^u du' \; (u-u') \; e^{-\frac{\gamma^2}2 \; (u-u')^2}
 \; \frac{\td(k,u')}{[1-u']^2} = I[\gamma \, u] \; ,
\ee
where $ I[\gamma \, u] $ [eq.(\ref{gorda})] is the inhomogeneity determined by the
initial condition. Taking two derivatives with respect to $ u $ and
using eq.(\ref{gileq})
we obtain
\be\label{diffinteq}
\ddot{\td}(k,u)  - \frac6{1-u^2} \; \td(k,u)+
3 \; \gamma^2 \; \td(k,u) - 6  \; \gamma^4 \; \int_0^u du' \; (u-u')^3 \;
e^{-\frac{\gamma^2}2 \; (u-u')^2} \; \frac{\td(k,u')}{[1-u']^2} =
\ddot{I}[\gamma \, u]+3 \; \gamma^2  \; I[\gamma \, u] \; .
\ee
From this expression we obtain the explicit form  of the kernel
$ K(u-u') $ that defines the source term eq.(\ref{S1}) and enters in the
transfer function eq.(\ref{Tint}), namely
\be\label{KMB}
K(u-u') = 6 \; \gamma^4  \; (u-u')^3 \; e^{-\frac{\gamma^2}2 \; (u-u')^2} \; .
\ee
In the cosmologically relevant case $ \gamma \leq 10^{-5} $,  we can
safely approximate the density perturbation, gravitational
potential and transfer function by their Born term in the Fredholm
series, namely
\bea \label{tdMB1}
&& \td^{(1)}(k,u) = I[\gamma \, u] + \frac6{\sqrt3 \; \gamma}
\int_0^u \left[h_1(u) \; h_2(u')-h_2(u) \; h_1(u')\right] \;
\frac{I[\gamma \, u']}{(1-u')^2} \;  du'   \; , \\
&&  \Phi^{(1)}(k,u) = (1-u)^2  \; \td^{(1)}(k,u)  \; ,\label{tfiMB1} \\
&& T_B(k)= \frac{10}{\sqrt3 \; \gamma^3} \int^1_0 h_2(u) \;
 \frac{I[\gamma \, u]}{(1-u)^2}  \; du \; . \label{TofkMB1}
\eea
 For the sake of comparison with the exact result and to display the
 dependence on initial conditions we study both initial conditions
eqs.(\ref{adiapert}) and (\ref{gilini}). The calculation of $ I[\gamma \, u] $ in
 both cases is best performed with cartesian coordinates, we find
 for the case of temperature perturbations eq.(\ref{adiapert})
\be \label{ITMB}
I_T[\gamma \, u]= \left[1-\frac13 \; \gamma^2  \; u^2\right] \;
e^{-\frac12 \; \gamma^2 \;  u^2} \; ,
 \ee
and for Gilbert's initial condition eq.(\ref{gilini})
\be\label{IGMB}
 I_G[\gamma \, u]=  e^{-\frac12 \; \gamma^2  \; u^2} \; .
\ee
They are connected by the simple relationship
\be\label{relsim}
I_T[\gamma \, u]= \left( 1 + \frac13 \; \gamma \;
\frac{\partial}{\partial \gamma} \right)  I_G[\gamma \, u] \; .
\ee

\begin{figure}[ht!]
\begin{center}
\includegraphics[height=2in,width=3in,keepaspectratio=true]{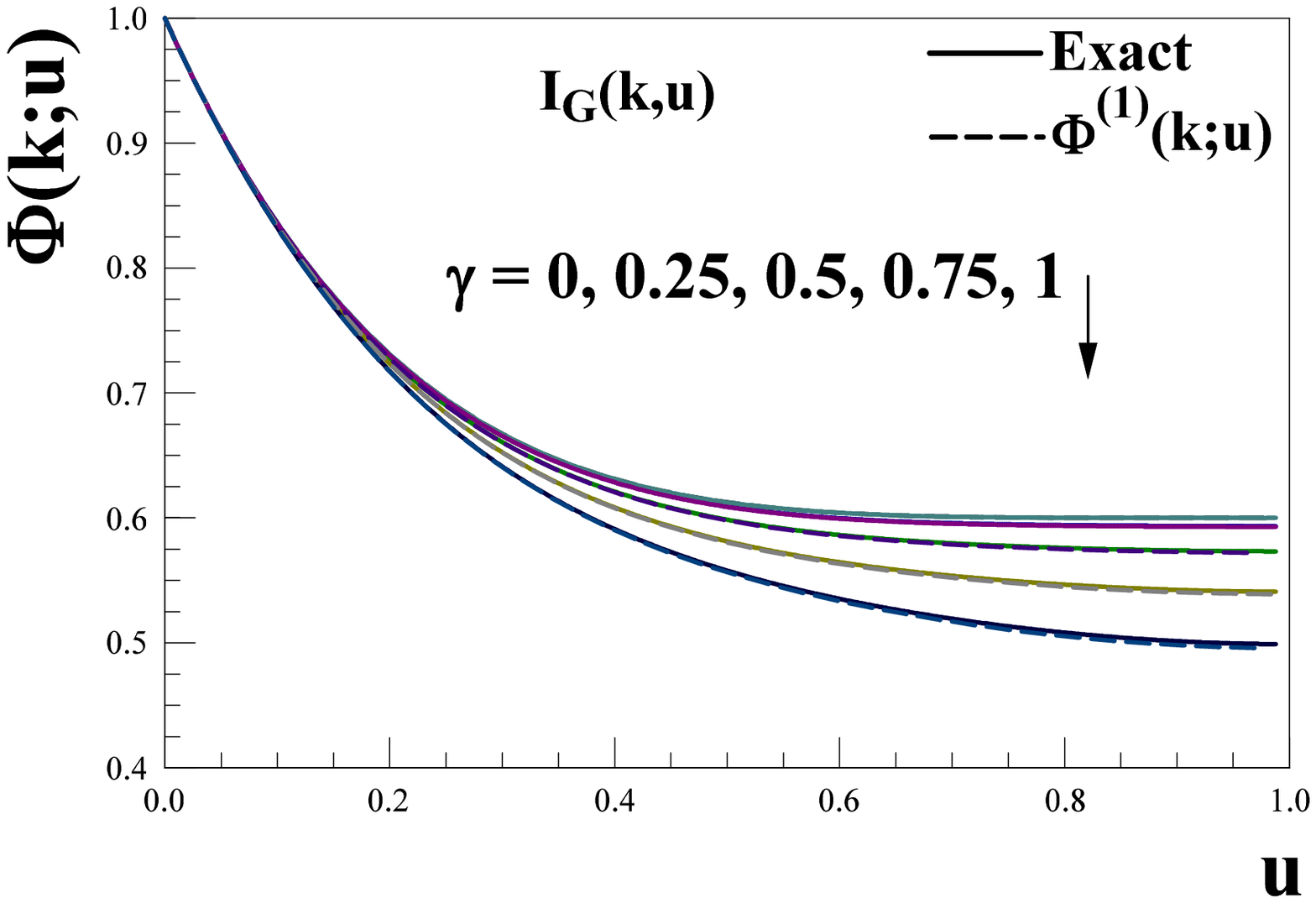}
\includegraphics[height=2in,width=3in,keepaspectratio=true]{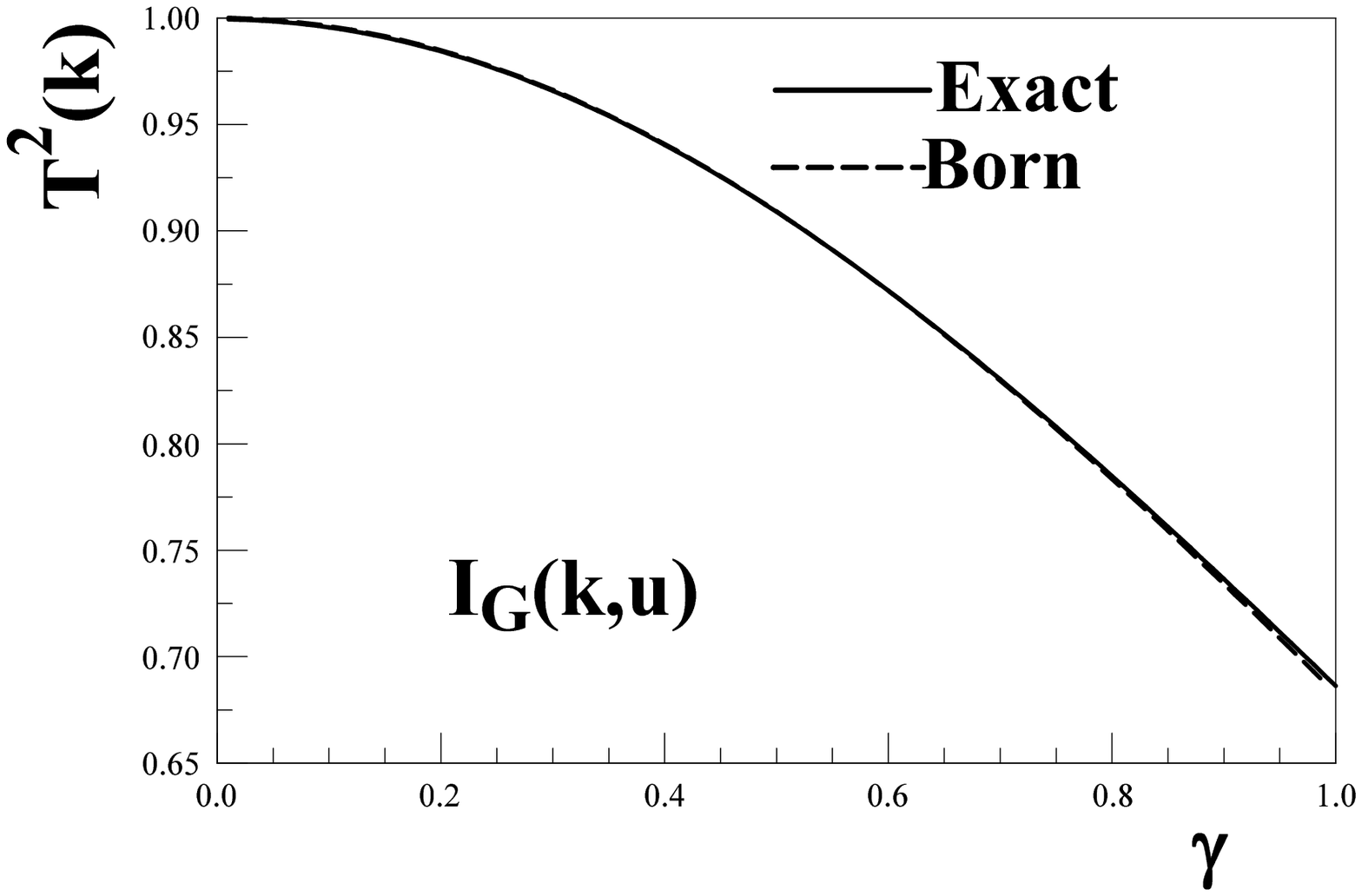}
\caption{Non-relativistic, Maxwell-Boltzmann distribution.
Left panel: The gravitational potential
$ \Phi(k,u) $ vs. $ u $ for
$ \gamma=0, \; 0.25, \; 0.4, \; 0.75, \; 1 $. Right panel: the transfer function
$ T^2(k) $ vs. $ \gamma $. In
both cases the solid line corresponds to the exact solution of eq.
(\ref{gil2}) with the initial condition eq.(\ref{IGMB}) and the dashed
line to the Born approximation eqs.(\ref{tfiMB1}) and (\ref{TofkMB1}).
} \label{fig:gilini}
\end{center}
\end{figure}

 \begin{figure}[ht]
\begin{center}
\includegraphics[height=2in,width=3in,keepaspectratio=true]{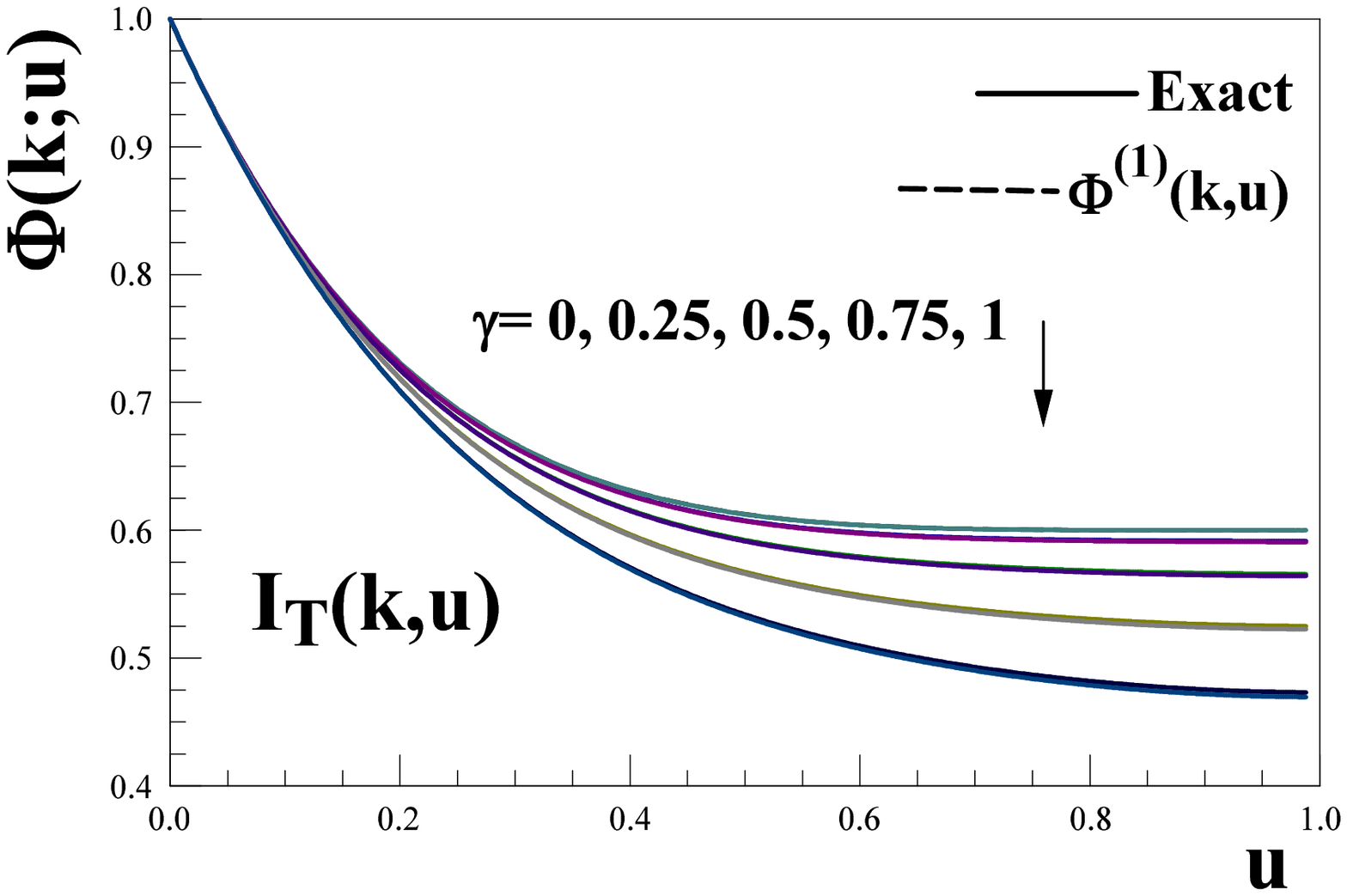}
\includegraphics[height=2in,width=3in,keepaspectratio=true]{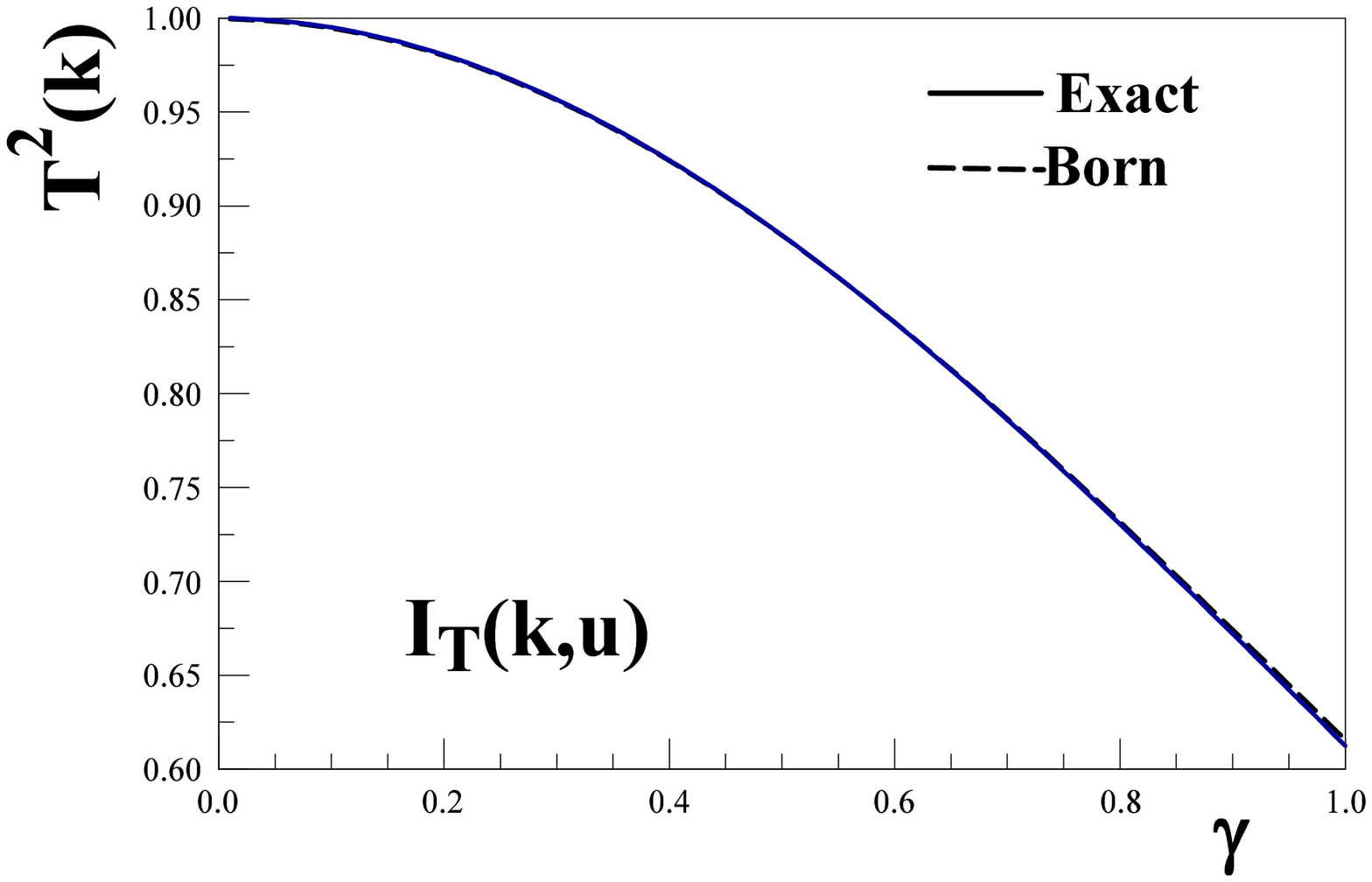}
\caption{Non-relativistic, Maxwell-Boltzmann distribution.
Left panel: $ \Phi(k,u) $ vs $ \gamma $ for
$ \gamma=0,0.25,0.4,0.75,1 $. Right panel: $ T^2(k) $ vs. $ \gamma $. In
both cases the solid line corresponds to the exact solution of eq.
(\ref{gil2}) with the initial condition eq.(\ref{ITMB}) and the dashed
line to the the Born approximation eqs.(\ref{tfiMB1}) and (\ref{TofkMB1}).
} \label{fig:temini}
\end{center}
\end{figure}

\begin{figure}[ht]
\begin{center}
\includegraphics[height=2in,width=3in,keepaspectratio=true]{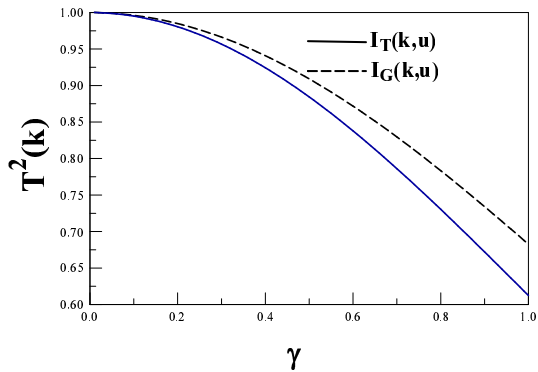}
\caption{Non-relativistic, Maxwell-Boltzmann distribution.
Comparison between $ T^2(k) $ vs. $ \gamma $ for the initial
conditions eqs.(\ref{ITMB}) and (\ref{IGMB}). The exact result and the
Born approximation are indistinguishable in this range.} \label{fig:compa}
\end{center}
\end{figure}

Figs. \ref{fig:gilini} and \ref{fig:temini} display the exact
results for $ \Phi(k;u) $ and $ T^2(k) $ obtained by numerical
integration of eq.(\ref{gileq}) and using the relation (\ref{fidelta2}),
compared to the solution obtained with
the Born approximation. It is clear that the Born
approximation is remarkably accurate for $ \gamma \lesssim 1 $, namely $k \lesssim k_{fs}(t_{eq})$.
Fig. \ref{fig:compa} compares $ T^2(k) $ for the initial conditions
eq.(\ref{IGMB}) and eq.(\ref{ITMB}). The exact numerical results are
indistinguishable from the Born approximation in the range
displayed.

\subsection{Short wavelength: memory of gravitational clustering}\label{MarkoMB}

Although for WIMPs only the region of $ \gamma \ll 1 $ is of
cosmological relevance as discussed above, it is important
  to study the opposite limit $ \gamma \gg 1 $ to
understand how the memory kernel $ K(u-u') $ modifies the transfer
function. It proves convenient to change variables to $ z=\gamma \;
(u-u') $ and to replace in eq.(\ref{S1}) the density perturbation
$ \delta $ by the gravitational potential $ \Phi(k,u) $ which is bounded
in time, to write
\be \label{Slk}
S_{NB}(u)= 6 \; \int_0^{\gamma\,u} z^3 \; e^{-\frac{z^2}2}  \;
\frac{\Phi\big[k;u-\frac{z}{\gamma}\big]}{\big[1-u+
\frac{z}{\gamma}\big]^4}~dz
\ee
Since $ \Phi(k;u\sim 0) \sim 1 $ it is clear from
this expression that $ S_{NB} $ is negligible either at small time
($ u\ll 1/\gamma $) or in the long wavelength limit $ \gamma \ll 1 $.
The factor $  z^3 \; e^{-\frac{z^2}2} $ in the
integrand grows as $ z^3$ at small $ z $, attains a maximum at $ z
=\sqrt3 $, namely $ u-u' \sim 1/\gamma$, and falls-off
exponentially for $ z > \sqrt3 $ for $ 1-u \gg 1/\gamma $.

Therefore, $ S_{NB} $ begins to contribute for $ u > 1/\gamma $.
Since $ \Phi(k,u) $ is still $ \sim \mathcal{O}(1) $ during this interval
it follows that $ S_{NB} $ becomes of $ \mathcal{O}(1) $,
whereas for large $ \gamma $ the Born term for Gilbert's initial
conditions,
\be\label{SBG}
S_{B,G} = \gamma^2  \; \left[2 + \gamma^2  \; u^2\right] \;
e^{-\frac12 \; \gamma^2 \;  u^2} \; ,
\ee
is suppressed for $ u \gg 1/\gamma $. Analogous conclusions follow for
temperature perturbations where the expression for $ S_{B,A} $ follows
combining eqs.(\ref{relsim}) and (\ref{SBG}).

\medskip

Therefore, for $ \gamma \gg 1 $ there is a crossover in behavior:
during $ 0\leq u \leq 1/\gamma $ the Born term which dominates is of
$ \mathcal{O}(1) $ and the second order correction is negligible, while at
$ u \sim 1/\gamma $ both terms become of the same order, and
for $ u > 1/\gamma $ the second order correction becomes important.
Finally, for $ 1-u \ll 1/\gamma $ the small $ z $
region dominates yielding a logarithmic behavior for the second order
correction. When this
correction becomes important for $ \gamma \gg 1 $, the transfer
function  $ T(k) $ is very small, suppressed at least by the prefactor
$ 1/\gamma^3 $.

\medskip

For $ \gamma \gg 1 $, in the region with $ \gamma u \gg 1 $ the upper
limit of the integral can
be taken to infinity. For $ 1/\gamma \ll u \ll 1-1/\gamma $  the
dominant contribution arises from the region $ u \sim 0 $ where $ \Phi
\sim 1 $ is largest, in this region the full integral is of
$ \mathcal{O}(1) $.

\medskip

Furthermore, since to leading order $ \Phi $ decays
via free-streaming on a time scale $ u \sim 1/\gamma $, its derivative
is large during the initial free-streaming regime but becomes small
for $ u \geq 1/\gamma $ as can be gleaned from Figs
\ref{fig:gilini}-\ref{fig:temini}. This analysis leads to the
conclusion that $ S_{NB}(u) $ becomes important for $ \gamma \gg 1 $   in
the region $ \gamma \; u \gg 1 $, where it is of the same order (or
larger) than the free streaming contribution (Born term).
When  $ S_{NB}(u) $ becomes non-negligible the transfer function is
suppressed by $ \sim 1/\gamma^3 $. Therefore, the corrections to the
Born term become relevant at small scales when the transfer function
has diminished substantially.

In the region $ u\gg 1/\gamma $ the following \emph{Markovian}
approximation is reliable:
\be \label{marko}
\Phi\big[k;u-\frac{z}{\gamma}\big]
\approx \Phi\big[k;u \big]- \dot{\Phi}\big[k;u \big] \;
\frac{z}{\gamma} +\cdots \; .
\ee
We emphasize that this Markovian approximation is valid only after the initial free
streaming transient for $ u > 1/\gamma $, since during the transient
$ \dot{\Phi}\sim \gamma $. Therefore, although such Markovian
approximation is available after the initial transient, the value of
the gravitational potential must be \emph{matched} to that at the
end of the free-streaming period in order to provide the full
dynamical evolution.

The analysis above suggests that the influence of the memory kernel
$ K(u-u') $ becomes important for short wavelengths $ \gamma >1 $. Thus,
for scales $ \lambda \leq\lambda_{fs}(t_{eq})\sim l_{fs}(v;0) $ where
$ \gamma > 1 $ we need to include the \emph{second order} term in the
Fredholm expansion in the transfer function given by
eq.(\ref{Tint2ndord}), where
\be
S_{NB}[\delta^{(1)};u] = 6  \; \gamma^4 \int_0^u du' \; (u-u')^3 \;
e^{-\frac{\gamma^2}2 \; (u-u')^2} \; \frac{\td^{(1)}(k,u')}{[1-u']^2}
\label{2ndordT}
\ee
and $ \td^{(1)}(k,u) $ is given by eq.(\ref{tdMB1}).
Higher order terms are further suppressed by extra
powers of $ 1/\gamma $ in the transfer function.

The addition of the second order correction yields a {\it remarkably
accurate} fit to the exact solution of the Boltzmann-Vlasov equation
in a wide range of scales down to scales  $ \lambda \ll
\lambda_{fs}(t_{eq}) $.

\begin{figure}[ht]
\begin{center}
\includegraphics[height=2in,width=3in,keepaspectratio=true]{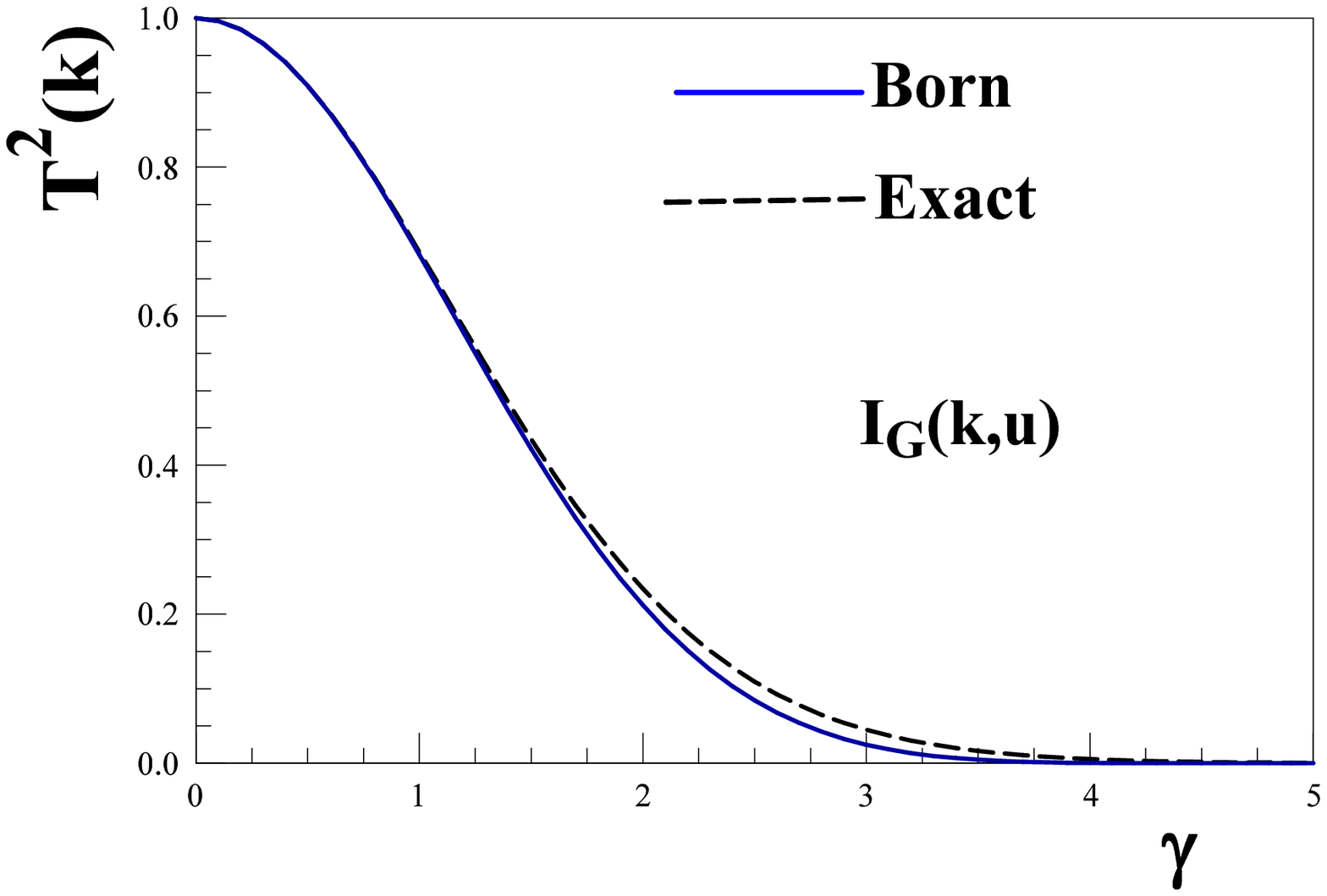}
\includegraphics[height=2in,width=3in,keepaspectratio=true]{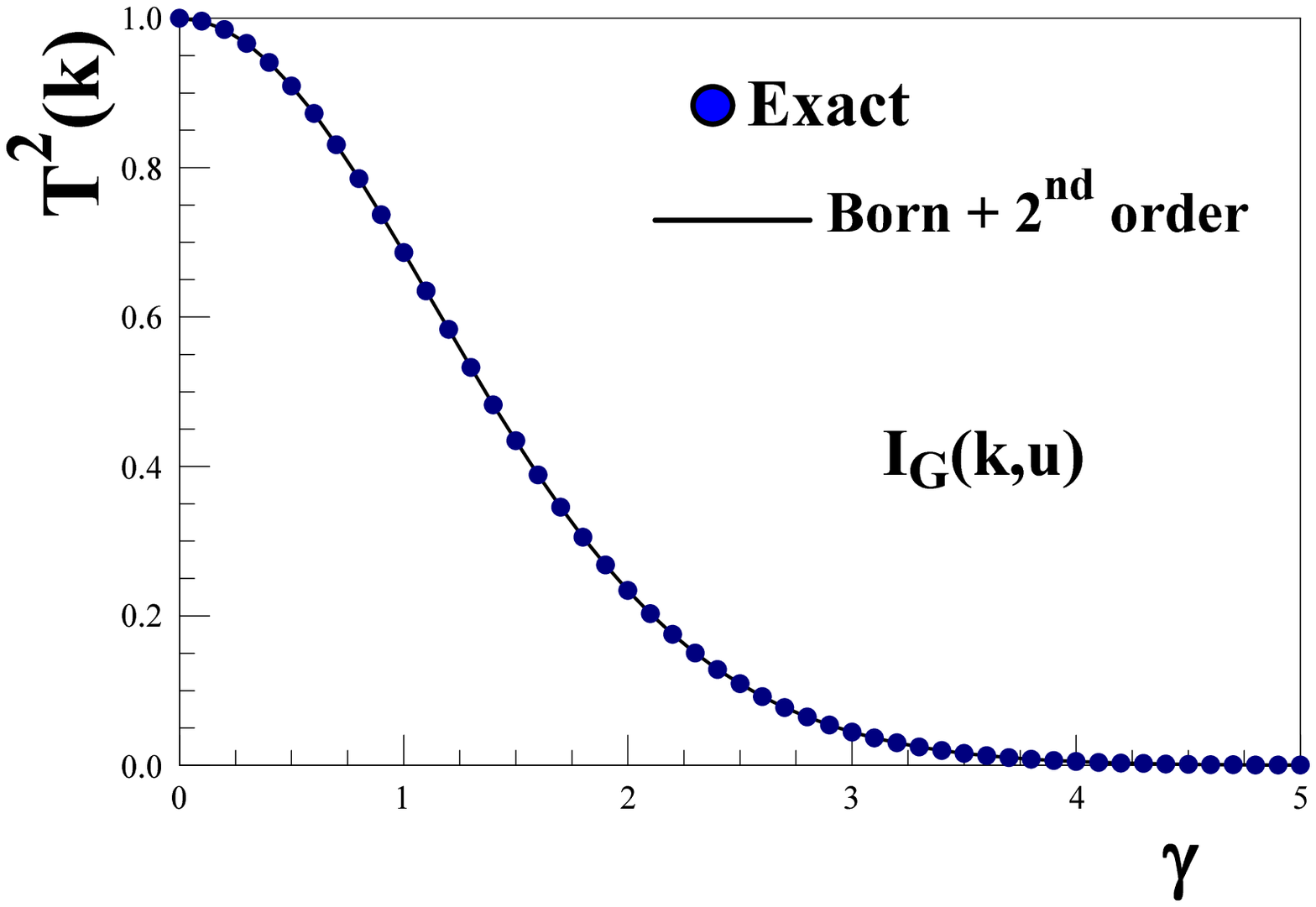}
\caption{Non-relativistic, Maxwell-Boltzmann distribution.
Left panel: The transfer function
$ T^2(k) $ vs. $ \gamma $ for the exact solution and
the Born approximation. Right panel: $ T^2(k) $ vs. $ \gamma $ for the
exact solution and the Born approximation plus second order.
Gilbert's initial condition eq.(\ref{IGMB}) was used.} \label{fig:mbtofk2}
\end{center}
\end{figure}

\begin{figure}[ht]
\begin{center}
\includegraphics[height=2in,width=3in,keepaspectratio=true]{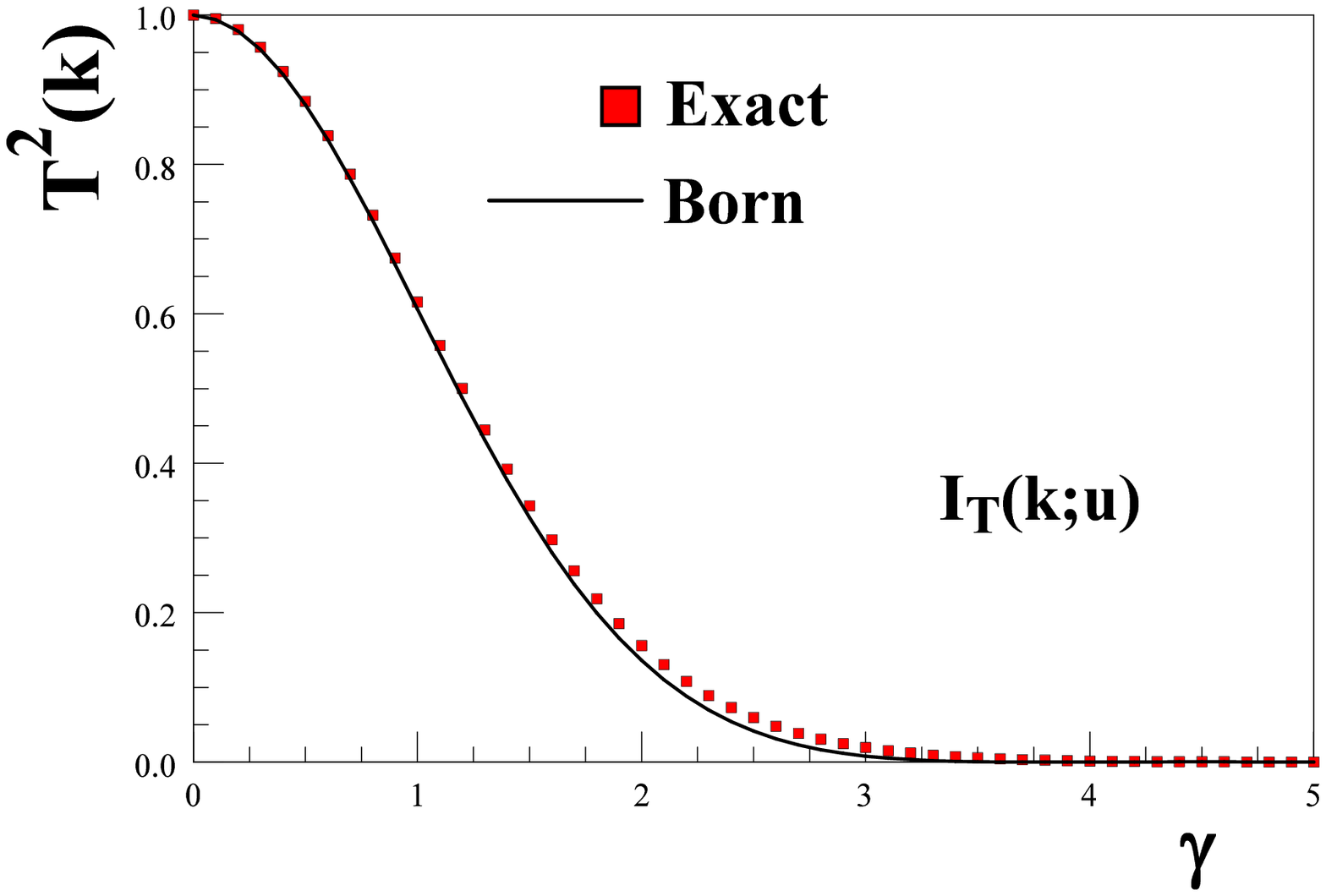}
\includegraphics[height=2in,width=3in,keepaspectratio=true]{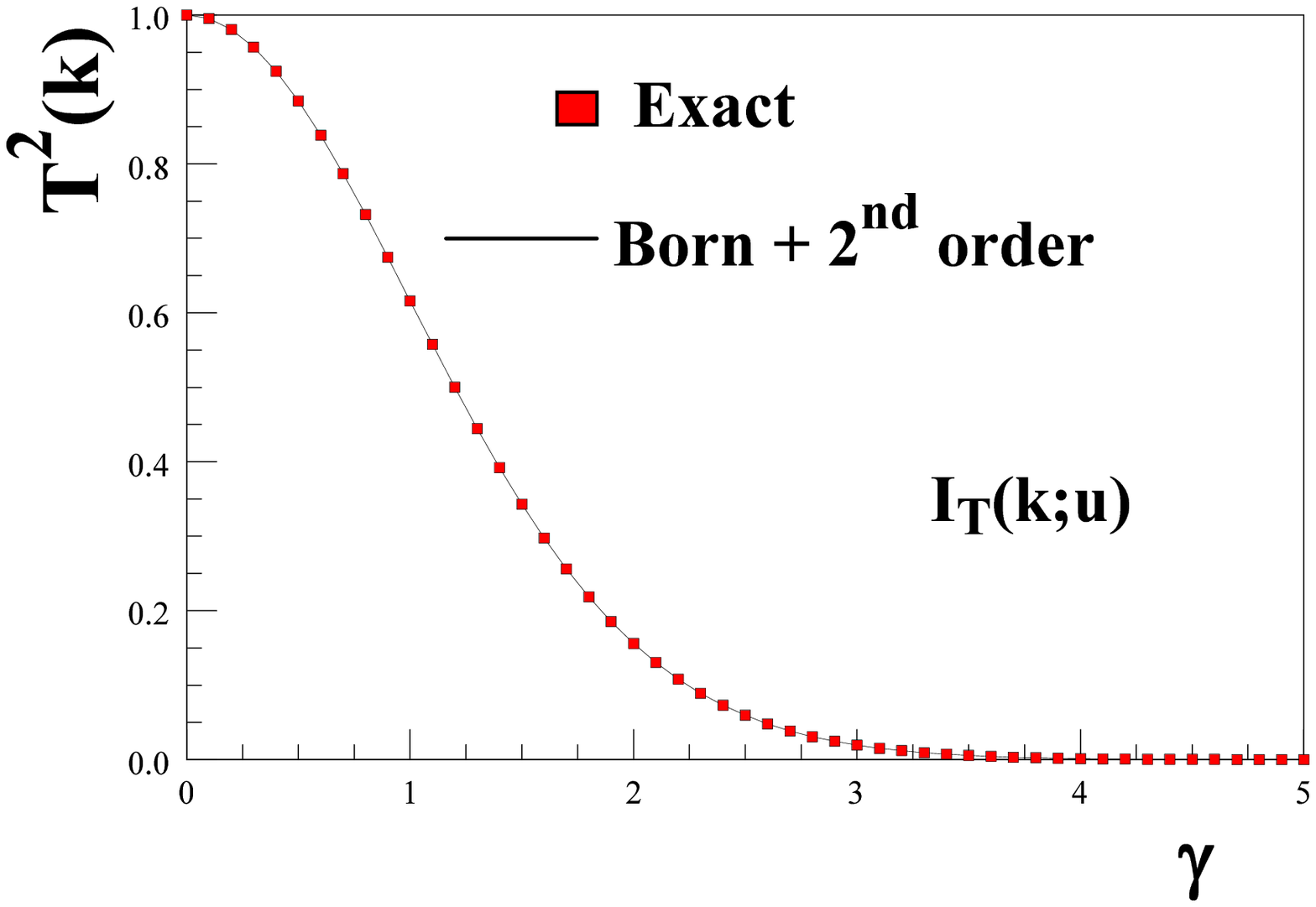}
\caption{Non-relativistic, Maxwell-Boltzmann distribution.
Left panel: the transfer function $ T^2(k) $ vs $ \gamma $ for the exact
solution and the Born approximation. Right panel: $ T^2(k) $ vs.
$ \gamma $ for the
exact solution and the Born approximation plus second order.
Initial condition for temperature fluctuations eq.(\ref{adiapert}) }
\label{fig:mbtofk2tini}
\end{center}
\end{figure}

Fig. \ref{fig:mbtofk2} displays the comparison
between the exact result for $ T^2(k) $ compared with both the Born
approximation and the next to Born approximation for initial
conditions eq.(\ref{IGMB}). Fig. \ref{fig:mbtofk2tini} displays
the same comparison for the initial condition  eq.(\ref{ITMB}). These
figures confirm the analysis presented above: the Born term is a
fairly accurate approximation in the region $ \gamma \lesssim 1 $ and
the second order correction becomes significant at $ \gamma \sim 1 $.
Including the second order correction yields a remarkably accurate
approximation in a wide range of scales $ 0 < \gamma < 5 $.

We obtain $ T(k) $ for large scales expanding eq.(\ref{IGMB}) for small $\gamma\,u$
\be \label{gamachi}
I_G[\gamma \, u]\buildrel{ \gamma \; u \to 0}\over= 1 - \frac12 \; \gamma^2 \; u^2
+ {\cal O}(\gamma^4 \; u^4) \; ,
\ee
and follow the same steps leading to  eq.(\ref{Talf2}),
  setting $ b= 1/2 $ according to eqs.(\ref{Iuchi}) and (\ref{gamachi}),  and replacing
$ \alpha $ by $ \gamma $ in eq.(\ref{Talf2}), with the result
\be\label{taluniv}
T(k) = 1 - \left( \frac{\gamma}{\gamma_0}\right)^2 + {\cal O}(\gamma^4)
\ee
where
\be \label{gamacero}
\gamma_0 = \left\{
\begin{array}{l}
\sqrt{\frac{21}{8}} = 1.62\ldots \quad {\rm Gilbert ~ initial ~ conditions,}
\\
\sqrt{\frac{126}{62}} = 1.42\ldots
\quad {\rm Temperature ~ initial ~ conditions.}
\end{array} \right.
\ee
$ \gamma_0 $ characterizes the fall-off of $ T(k) $. We find for a given initial condition $\gamma_0$ is
independent  of the statistics for thermal relics.

For WIMPs the long-wavelength approximation (\ref{taluniv}) describes $T(k)$ in the whole range of
scales of cosmological relevance for structure formation in the linear regime, since for these scales $\gamma \lesssim 10^{-5}$
and the Born approximation describes the DM density and gravitational perturbations outstandingly well.

Although for the scales of interest for structure formation $T(k)$ given by (\ref{taluniv}) gives the correct description, it is
illuminating to study the small scale behavior of $T(k)$.

The precise numerical  solution of eq.(\ref{gileq}) shows that
$ T(k) $ for temperature initial conditions
has a zero at $ \gamma = 7.7 \ldots $ as shown
in fig. \ref{fig:gamagrande} and is negative for
$ \gamma >7.7\ldots $     decreasing
exponentially for $ \gamma > 10 $  as
\be\label{fitMB}
 |T(k)| \simeq e^{-\left(\frac{\gamma}{\gamma_{MB}}\right)^{x_{MB}}}
\ee
where $ x_{MB} \simeq 1.3 $ and $ \gamma_{MB} \simeq 1.55 $ both for
temperature and Gilbert initial conditions eqs.(\ref{adiapert}) and
(\ref{gilini}).

\medskip

We draw the following important lessons from this study:

\begin{itemize}
\item{For $ k \leq k_{fs}(t_{eq}) $,
the Born approximation gives a simple and very accurate description
of the transfer function. It inputs the
free streaming solution in absence of self-gravity and the mode
functions associated with the fluid description. These \emph{only}
input the average squared velocity, therefore the first moment
$\langle p^2 \rangle$ of the distribution function. This approximate
but fairly accurate solution yields a simple tool to understand
different distribution functions and initial conditions in generic
situations.  }

\item{The short wavelength region of the transfer function
$ k > k_{fs}(t_{eq})$ is remarkably well described by
correcting the Born approximation with the second order term in the
Fredholm solution. This correction incorporates higher moments of the
distribution function and includes important {\it memory effects} of
gravitational clustering and corrections to the fluid description. This approximation while just slightly more
involved than the Born term also yields a rather simple and systematic
tool to study the effects of {\it higher order correlations} and the full
structure of the distribution function along with memory
of gravitational clustering in generic situations.  }

\item{We see that the scale characterizing the suppression of $ T(k) $
with $ \gamma $ {\bf decreases} with $ k $ (and $ \gamma $).
Namely, $ \gamma_0 $ characterizing the suppression for small $ k $
is larger than the characteristic suppression scale for larger $ k $.

Nevertheless, the numerical study shows that the characteristic suppression scale $ k_{char} $
related to a characteristic dimensionless ratio $ \gamma_{char} $
by eqs.(\ref{T0d}), (\ref{alfa}), (\ref{alfanum})  and (\ref{overy2MB}), namely:
\be \label{kcarMB}
k_{char} = \frac12 \;  H_0  \; \sqrt{\Omega_M \; a_{eq}} \; \left(\frac{g_d}2\right)^\frac13
\; \frac{\sqrt{T_d \; m}}{T_{cmb}} \; \gamma_{char} =
4.17 \;  \left(\frac{g_d}2\right)^\frac13 \; \Big(\frac{m}{100 \,\mathrm{GeV}}
\Big)^\frac12 \; \Big(\frac{T_d}{10 \,\mathrm{MeV}} \Big)^\frac12 \;
\frac{\gamma_{char}}{\mathrm{pc}} \; .
\ee
is such that in all cases considered we find   $ \gamma_{char} \sim \mathcal{O}(1)$. Therefore,
$ k_{fs}(t_{eq}) $ obtained from  free particle propagation [eqs.(\ref{lfsv})-(\ref{equiva})] and
  given by equations (\ref{kofs},\ref{kofsCDM}) only differs by a factor
of order one from $ k_{char} $ [eq.(\ref{kcarMB})].
  }

\end{itemize}

\section{Warm and hot dark matter: Fermions vs. Bosons}

In this section we consider thermal relics that decoupled in LTE while
ultrarelativistic, but
that have become non-relativistic during matter domination.
Their normalized distribution function after freeze-out are given by
\bea
\tilde{f}_0(y) &= &  \frac2{3 \; \zeta(3)} \;
\frac1{e^y+1} \quad \mathrm{FD} \label{FD} \; ,
\\ \tilde{f}_0(y) &= &  \frac1{2 \; \zeta(3)} \; \frac1{e^y-1}
\quad \mathrm{BE} \label{BE} \; , \quad  y = \frac{p}{T_{d,0}} \; ,
\eea
for FD and BE respectively, with
$ T_{d,0} $ being the decoupling temperature today. We study each case separately.

It is convenient to state the following general result in order to
estimate the large momentum contribution of the various integrals.
Consider a generic integral of the form
\be \label{Fint}
\int_0^\infty F(y) \;  \sin[yz] \; dy  \;  ,
\ee
where $ F(0), \; F'(0), \;F''(0) \cdots $ are
finite and $ F(\infty)=0 $. Integrating by parts consecutively we find the large $ z $ behavior
\be  \label{asyint}
\int_0^\infty F(y) \; \sin[yz] \; dy = \frac{F[0]}{z}-\frac{F''[0]}{z^3} +
\mathcal{O}\left(\frac1{z^5}\right) \; .
\ee
This result will be useful in the analysis that follows.

\bigskip

$ T(k) $ decreases in $ k $ with a characteristic scale $ k_{char} $
which depends on the initial conditions, the particle statistics and the
regime of wavenumbers. $ k_{char} $ translates into a characteristic scale
$ \alpha_{char} $ in the dimensionless variable $ \alpha $. $ k_{char} $
is related to $ \alpha_{char} $ by eqs.(\ref{T0d}), (\ref{alfa}) and
(\ref{alfanum}):
\be \label{kcar}
k_{char} = \frac12 \;  H_0  \; \sqrt{\Omega_M \; a_{eq}} \; \left(\frac{g_d}2\right)^\frac13
\; \frac{m}{T_{cmb}} \; \alpha_{char} = 0.00417 \; \alpha_{char} \; \left(\frac{g_d}2\right)^\frac13
\; \frac{m}{\mathrm{keV}} \; [\mathrm{kpc}]^{-1} \; .
\ee
For fermionic and bosonic thermal relics $ \alpha_{char} \sim \mathrm{O}(1)$ because in these
cases $\overline{y^2} \sim \mathrm{O}(1)$.

\subsection{Fermions}\label{sfer}

For fermions that decoupled relativistically in LTE with the
normalized distribution function eq.(\ref{FD}), it follows that
\be \label{overYF}
\overline{y^2} = \int^\infty_0 y^4  \; \tilde{f}_0(y)  \; dy  = 15 \;
\frac{\zeta(5)}{\zeta(3)} = 12.939\cdots \; ,
\ee
leading from eq.(\ref{kofs}) to the free-streaming wavevector today
\be  \label{kofsFD}
k_{fs}(t_{eq}) =  0.00284\cdots \, \Big(
\frac{g_d}2\Big)^\frac13 \; \frac{m}{\mathrm{keV}} \;
[\mathrm{kpc}]^{-1} \; .
\ee

For the initial condition eq.(\ref{adiapert}) the inhomogeneity is
\be \label{ITFD}
I_T[\alpha  \, u] = \frac2{9 \; \zeta(3)}\, \int^\infty_0
\frac{y^2 \; e^y}{(e^y+1)^2} \; \frac{\sin[y \; \alpha  \; u]}{\alpha \;  u} \;
dy = \frac4{3 \; \zeta(3)} \sum_{n=1}^\infty
\frac{(-1)^{n+1} \; n}{(n^2+z^2)^2}\Big[1-\frac43 \; \frac{z^2}{(n^2+z^2)} \Big] \quad , \quad
z=\alpha \; u \; ,
\ee
whereas for the initial condition eq.(\ref{gilini})
 \be \label{IGFD}
I_G[\alpha  \, u] = \frac2{3 \; \zeta(3)}\int^\infty_0 \frac{y}{e^y+1} \;
  \frac{\sin[y  \; \alpha  \; u]}{\alpha  \; u} \;  dy = \frac4{3 \; \zeta(3)}
  \sum_{n=1}^\infty \frac{(-1)^{n+1} \; n}{(n^2+z^2)^2}\quad , \quad
z=\alpha  \; u \; .
\ee
Using the result eq.(\ref{asyint}) we find that for both initial conditions
the asymptotic behavior of the inhomogeneity for $ \alpha \; u \gg 1 $  is given by
\be \label{asyIFD}
I_G[\alpha  \, u] \buildrel{ \alpha \; u \to \infty}\over= \frac1{3\; \zeta(3) \;  (\alpha \; u)^4}
 +\mathcal{O}\Bigg(\frac1{[\alpha \;  u]^6}\Bigg) \quad ,  \quad
I_T[\alpha  \, u] \buildrel{ \alpha \; u \to \infty}\over= -\frac1{9\; \zeta(3) \;  (\alpha \; u)^4}
 +\mathcal{O}\Bigg(\frac1{[\alpha \;  u]^6}\Bigg) \; ,
\ee
which must be contrasted to the Maxwell-Boltzmann case for which
$ I_G[\alpha  \, u] $ and $ I_T[\alpha  \, u] $ eqs.(\ref{ITMB}) and
(\ref{IGMB}) decay exponentially.

Therefore, FD statistics results in free-streaming solutions in absence
of self-gravity $ I_G[\alpha  \, u] $ and $ I_T[\alpha  \, u] $
that fall off much slower, with a power law in time.
Such long-range feature is also present in the kernel $ K(u-u') $.
This is unlike the Maxwell-Boltzmann case where such free-streaming
solutions eqs.(\ref{ITMB}) and (\ref{IGMB})
fall-off exponentially on a time scale $ 1/\gamma $.

\medskip

\begin{figure}[h]
\begin{center}
\includegraphics[height=2in,width=3in,keepaspectratio=true]{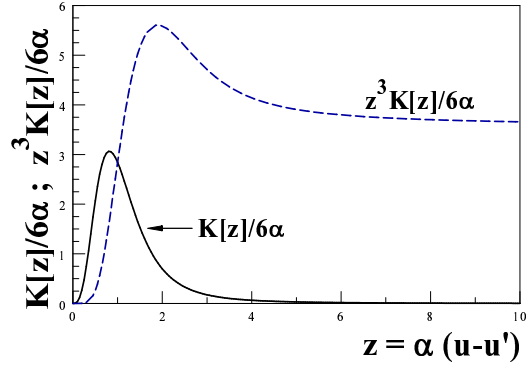}
\caption{The kernel $ K(u-u')/[6 \; \alpha] $ [see eq.(\ref{kernel})]
vs. $ z= \alpha \; (u-u') $ for Fermionic thermal relics.}
\label{fig:kernelfermion}
\end{center}
\end{figure}

Implementing the result eq.(\ref{asyint}) we find  that the kernel
$ K(u-u') $ eq.(\ref{kernel}) with the normalized FD
distribution function eq.(\ref{FD}) falls off as
\be \label{Kfasi}
K(u-u')\buildrel{ \alpha \; (u-u') \to \infty}\over=\frac{30 \; \zeta(5)}{\zeta^2(3)}
\; \frac{\alpha}{\left[\alpha \; (u-u')\right]^3} \quad {\rm FD} \quad \; .
\ee
Fig. \ref{fig:kernelfermion} displays both $ K[z]/[6 \; \alpha] $ and
$ z^3 \; K[z]/[6 \; \alpha] $. Again this case must be contrasted with
the Maxwell-Boltzmann case eq.(\ref{KMB}) which decays exponentially.

Thus, whereas for $ \gamma \gg 1 $ the Maxwell-Boltzmann distribution
leads to a short range memory kernel, the FD distribution yields
a long range kernel that keeps memory of the initial state and the
initial value of the gravitational and density perturbations.

For the FD case eqs.(\ref{gamma}) and (\ref{overYF}) yield
\be \label{alfagaFD}
\alpha = \sqrt{\frac{\zeta(3)}{5 \; \zeta(5)}} \; \gamma
= 0.482 \ldots \,\gamma \; .
\ee
We can now study the transfer
function as a function of $ \gamma $ comparing the exact result with
the Born approximation eq.(\ref{Tborn}) and the next to Born
term eq.(\ref{T2ndord}). The analysis presented in the previous
section of the different contributions also applies to this case:

\medskip

For long wavelengths $ \gamma <1 $ the second order correction
 $S[\delta^{1},u] $ is much smaller than the Born term because the
kernel $ K(u-u') $ is of order $ \gamma^4 $, hence $ T(k) $ is dominated
by the Born term and by the free streaming solution.

\medskip

At shorter
wavelengths $ \gamma >1 $ the kernel $ K(u-u') $ begins to contribute at
$ u > 1/\gamma $ and peaks at $ u'\sim u $ as indicated by Fig.
\ref{fig:kernelfermion} but by this time both the density
perturbation and gravitational potential inside the integrand have
decayed substantially leading to a suppressed contribution to the
transfer function.

This analysis leads to conclude that for $ \gamma
< 1, \; T(k) $ is dominated by the Born term while second (and higher)
order corrections become relevant for $ \gamma >1 $.

\medskip

\begin{figure}[ht]
\begin{center}
\includegraphics[height=2in,width=3in,keepaspectratio=true]{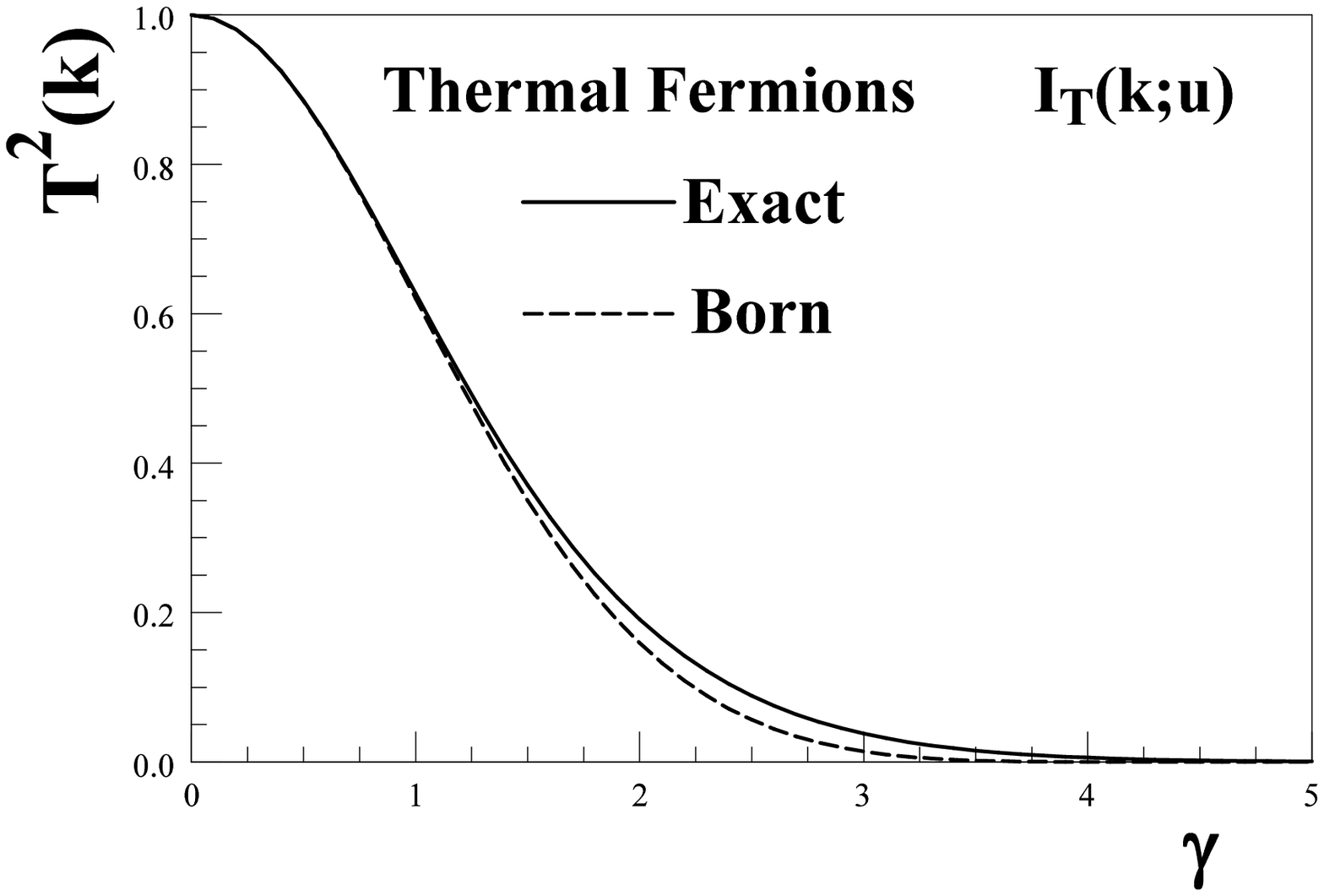}
\includegraphics[height=2in,width=3in,keepaspectratio=true]{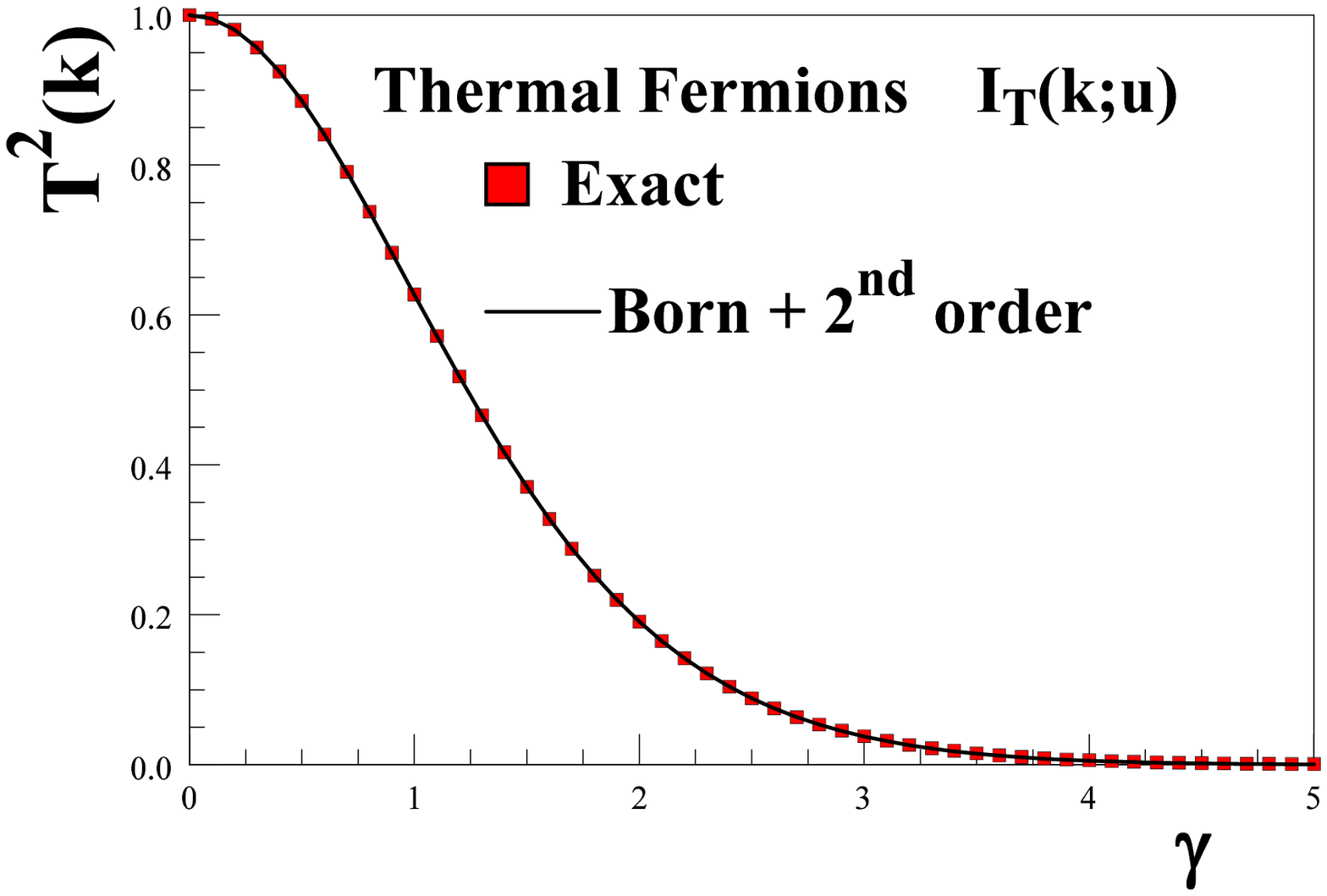}
\caption{Fermionic thermal relics.
Left panel: $ T^2(k) $ vs $ \gamma $ for the exact solution and
the Born approximation. Right panel: $ T^2(k )$ vs. $ \gamma $ for the
exact solution and the Born approximation plus second order.
Initial condition given by eq.(\ref{ITFD}). Similar agreement is found
with the initial Gilbert's conditions eq.(\ref{IGFD}). }
\label{fig:thermalfermionIT}
\end{center}
\end{figure}

We have confirmed this analysis by comparing the exact numerical solution
of Gilbert's equation with the Born approximation and the second order
correction to the transfer function for both initial conditions
eqs.(\ref{ITFD}) and (\ref{IGFD}). The results are qualitatively the same
in both cases and we present them for temperature initial conditions
in fig. \ref{fig:thermalfermionIT}. The agreement between the exact result
and the Born plus second order correction is remarkable. The
numerical study confirms the analysis above,  the second order
contribution, which includes the memory of gravitational clustering
becomes important  for  $ \gamma >1 $, namely at small scales,
but when it begins to be appreciable,  $ T(k) $ has been highly
suppressed. For $ \gamma \gtrsim 5 $ we find that
$ T^2(k) \lesssim 10^{-5} $.

\medskip

$ T(k) $ for large scales follows from eq.(\ref{Talf2}) and
$$
I_G[\alpha  \, u] \buildrel{ \alpha \; u \to 0}\over= 1 - \frac52 \;
\frac{\zeta(5)}{\zeta(3)} \; (\alpha \; u)^2
+ {\cal O}\left([\alpha \; u]^4\right) \; ,
$$
for fermions. We find that the large scale approximation for $ T(k) $
eq.(\ref{taluniv}) is {\bf also valid} in the FD case.

\medskip

The precise numerical resolution of eq.(\ref{gileq}) for fermions shows
that $ T(k) $ for Gilbert initial conditions  [eq.(\ref{gilini})]
and $ \gamma > 4 $ decreases as
\be\label{fitfg}
T(k) \simeq \left(\frac{\gamma_{fg}}{\gamma}\right)^{x_{fg}} \quad , \quad
 x_{fg} \simeq 7.6  \quad , \quad\gamma_{fg} \simeq 3.7 \; ,
\ee
see fig. \ref{fig:gamagrande}.

\medskip

The behavior of $ T(k) $ for temperature initial conditions
[eq.(\ref{adiapert})] is more involved as displayed in fig. \ref{fig:gamagrande}:
we find that $ T(k) $ decreases exponentially for $ 4 < \gamma < 15 $ as
\be\label{feradia}
T(k) \simeq c_{fa} \; e^{-\frac{\gamma}{\gamma_{fa1}}}
\quad , \quad  c_{fa} \sim 7.6 \quad , \quad \gamma_{fa1} \simeq 0.89 \; .
\ee
For $ \gamma > 15 $ $ T(k) $ decreases faster than in eq.(\ref{feradia}),
it vanishes and becomes negative at $ \gamma \simeq 16.72 $.
For $ \gamma > 25 $  $ T(k) $ decreases in
absolute value  as power law:
\be\label{fa2}
 T(k) \simeq -\left(\frac{\gamma_{fa2}}{\gamma}\right)^{x_{fa}}
\quad , \quad  x_{fa} \simeq 5.3  \quad , \quad \gamma_{fa2} \simeq 0.96
\; .
\ee

\medskip

We find that the scale of suppression   itself slides with scale.

For Gilbert initial conditions, we see that the characteristic fall-off
scale of $ T(k) $ {\bf increases} with increasing $ k $,   from $ \gamma_0
\simeq 1.62 \ldots $ for small $ k $ to $ \gamma_{fg} \simeq 3.7 $
for $ \gamma > 6 $. Instead, for temperature initial conditions,
the characteristic fall-off scale {\bf decreases} with increasing $ k $
reaching values $ \gamma_{fa} \simeq 0.89-0.96 $ for $ \gamma > 4 $. However, all of these are of the same order $\sim \mathcal{O}(1)$ which is  a manifestation of a \emph{unique characteristic} scale $\gamma_{char} \sim \mathcal{O}(1)$ as also found in the case of the Maxwell-Boltzmann distribution.

 Therefore, in terms of wavevector $k$ this observation translates into
the statement that the relevant scale for suppression of $T(k)$ is $k_{fs}(t_{eq})$, although the functional form of $T(k)$ itself depends on scale and initial
condition, in this
case varying from exponential to power law.

\medskip

We anticipate that the power law behavior of the kernel
$ K(u-u') $ for fermions [eq.(\ref{Kfasi})]
leads to a longer memory on the initial conditions than in
the Maxwell-Boltzmann case eq.(\ref{KMB}). For $ \gamma \gg 1 $  the range of the
Born term and that of the kernel are much longer for fermions
than for the Maxwell-Boltzmann case. Then, both the free streaming
solution and the gravitational perturbation (or alternatively the
density perturbation) for \emph{small} values of $ u' $ inside the
integrand in $ K(u-u') $ yield larger contributions for FD than for Maxwell-Boltzmann.
We indeed find that $ |T(k)| $ for thermal Fermions
and for a given value of $ \gamma $ is \emph{enhanced} relative to that of the
Maxwell-Boltzmann particles [see fig. \ref{fig:gamagrande}].

\subsection{Bosons}\label{sbos}

For thermal bosons that decoupled while relativistic with the normalized distribution function
eq.(\ref{BE}) we find
\be \label{overYB}
\overline{y^2} = 12 \; \frac{\zeta(5)}{\zeta(3)} = 10.352\cdots
\ee
leading from eq.(\ref{kofs}) to the free streaming wave vector today
\be  \label{kofsBE}
k_{fs}(t_{eq}) =  0.00317 \ldots \; \Big(
\frac{g_d}{2}\Big)^\frac13 \; \frac{m}{\mathrm{keV}} \;
[\mathrm{kpc}]^{-1} \; .
\ee

For the initial condition eq.(\ref{adiapert}) the inhomogeneity is
\be\label{ITBE}
I_T[\alpha  \, u] = \frac1{6 \; \zeta(3)}\,\int^\infty_0  \frac{
y^2 \; e^y}{(e^y+1)^2} \; \frac{\sin[y  \; \alpha  \; u]}{\alpha \;  u} \;  dy =
\frac1{\zeta(3)}  \; \sum_{n=1}^\infty \frac{
n}{(n^2+z^2)^2}\Big[1-\frac43\frac{z^2}{(n^2+z^2)} \Big] \quad , \quad
z=\alpha  \; u
\ee
and for the initial condition eq.(\ref{gilini})
\be \label{IGBE}
I_G[\alpha  \, u] = \frac1{2 \, \zeta(3)}\int^\infty_0
\frac{y}{e^y+1} \; \frac{\sin[y  \; \alpha  \; u]}{\alpha  \; u} \;  dy =
\frac1{\zeta(3)} \;
 \sum_{n=1}^\infty \frac{n}{(n^2+z^2)^2}\quad , \quad z=\alpha  \; u  \; .
\ee
Using the result eq.(\ref{asyint}) we find that the asymptotic behavior
of the inhomogeneity for $ \alpha  \; u \gg 1$ for both initial
conditions  is given by
\be \label{asyIBE}
I_G[\alpha  \, u] \buildrel{ \alpha \; u \to \infty}\over= \frac1{2 \; \zeta(3) \;  (\alpha \; u)^2}
 +\mathcal{O}\Bigg(\frac{1}{[\alpha\; u]^4}\Bigg) \quad ,  \quad
I_T[\alpha  \, u] \buildrel{ \alpha \; u \to \infty}\over= -\frac1{6 \; \zeta(3) \;  (\alpha \; u)^2}
 +\mathcal{O}\Bigg(\frac{1}{[\alpha\; u]^4}\Bigg) \; .
\ee
This is an even slower power law fall off than in
the FD case and even much slower than the Gaussian-exponential
fall off in the Maxwell-Boltzmann case.

\medskip

Since $ I[\alpha \, u] $ is the free streaming solution in absence of self-gravity
(normalized to one at the initial time), we see that with either
initial condition the Bose-Einstein distribution function leads to a
much less efficient free-streaming smoothing of the initial
perturbation.

\medskip

\begin{figure}[h]
\begin{center}
\includegraphics[height=2in,width=3in,keepaspectratio=true]{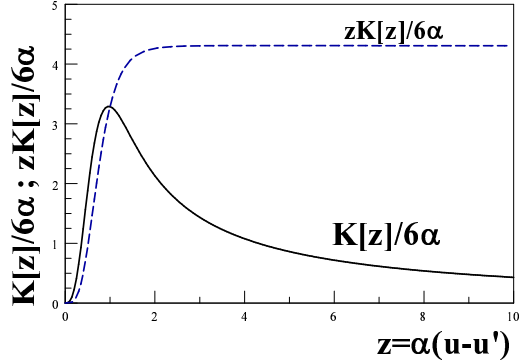}
\caption{The kernel $ K(u-u')/[6 \; \alpha] $ (full line) [see eq.(\ref{kernel})]
and $ z \; K[z]/(6 \; \alpha)$ (dashed line) vs.
$ z=\alpha \; (u-u') $ for Bosonic thermal relics. } \label{fig:kernelboson}
\end{center}
\end{figure}

This long range feature associated with the
Bose-Einstein distribution is also manifest in the kernel $ K(u-u') $.
Again, using eq.(\ref{asyint}) we find that the kernel $ K(u-u' )$ eq.
(\ref{kernel}) with the normalized Bose-Einstein distribution eq.(\ref{BE}) falls off as
\be \label{Kbasi}
K(u-u')\buildrel{ \alpha \; (u-u') \to \infty}\over=\frac{36 \; \zeta(5)}{\zeta^2(3)}
\; \frac1{u-u'} \quad {\rm BE} \quad \; .
\ee
Fig. \ref{fig:kernelboson} displays both $ K[z]/(6 \; \alpha) $ and
$ z \; K[z]/(6 \; \alpha)$. This fall-off is even slower than in the
FD and certainly much slower than the
Maxwell-Boltzmann case eq.(\ref{KMB}), with important consequences
explored below.

BE statistics, more specifically the behavior of the
distribution
function at small momentum, leads to a slow fall off with a power
law and \emph{long range memory}, even much longer than the FD
case because of the divergence of the distribution function as
$ y\rightarrow 0 $.

\medskip

The long-range nature of the kernel brings about
important consequences: even for  $u\sim 1$ the kernel is sensitive
to the region $ u' \sim 0 $, therefore the initial value of the
gravitational potential $ \Phi(k,u\sim 0) \sim 1 $, which is the
largest value that $ \Phi $ attains, contributes with a large measure
to the integrand. This feature in turn results in that
free-streaming is less efficient, and a large memory of the
initial value of the gravitational potential remains. This produces
an \emph{enhancement} in the transfer function as compared to the
Maxwell-Boltzmann and Fermi-Dirac cases, in which the memory in the
kernel is of much shorter range and the initial (maximum) value of
the gravitational potential does not contribute as much to the
integrand for $u \sim 1$.

This remarkable difference will be studied explicitly below in a
comparison between the three cases.

For bosons eqs.(\ref{gamma}) and (\ref{overYB}) yield,
\be \label{alfaBE}
\alpha = \sqrt{\frac{\zeta(3)}{4 \; \zeta(5)}} \; \gamma =0.538\ldots \,\gamma \; ,
\ee
and we now have
all the ingredients to study the transfer function as a function of
$ \gamma $ to compare to the previous cases.

\medskip

The analysis presented in
the previous cases for the magnitude of the contributions from the
Born and second (and higher) order remains the same. In the long
wavelength limit $ \gamma \ll 1 $ the Born term dominates, and the
second order correction is subleading of order $ \mathcal{O}(\gamma^4) $.

For large $ \gamma $ (small scales) just as in the Maxwell-Boltzmann
and Fermi-Dirac statistics, and quite generally, for $ u\sim 0 $ the
integral of the kernel yields a contribution $ \sim \gamma^4 \; u^4 $,
therefore for large $ \gamma $ the second order contribution begins to
be significant at $ u \sim 1/\gamma$  but at this time the
gravitational potential has decayed significantly for $ \gamma \gg 1 $.
Although the long range memory of the kernel maintains information
on the initial values of the gravitational potential, but suppressed
by a power $ 1/\gamma $. Therefore the second order correction becomes
significant for large $ \gamma $ (small scales) but is also suppressed
by inverse powers of $ \gamma $, and hence is always perturbatively small.

\medskip

It must be noticed that because of its longer range, the second order
correction for bosons is
comparatively \emph{larger} to that in the Maxwell-Boltzmann and
Fermi-Dirac case.

\medskip

\begin{figure}[ht]
\begin{center}
\includegraphics[height=2in,width=3in,keepaspectratio=true]{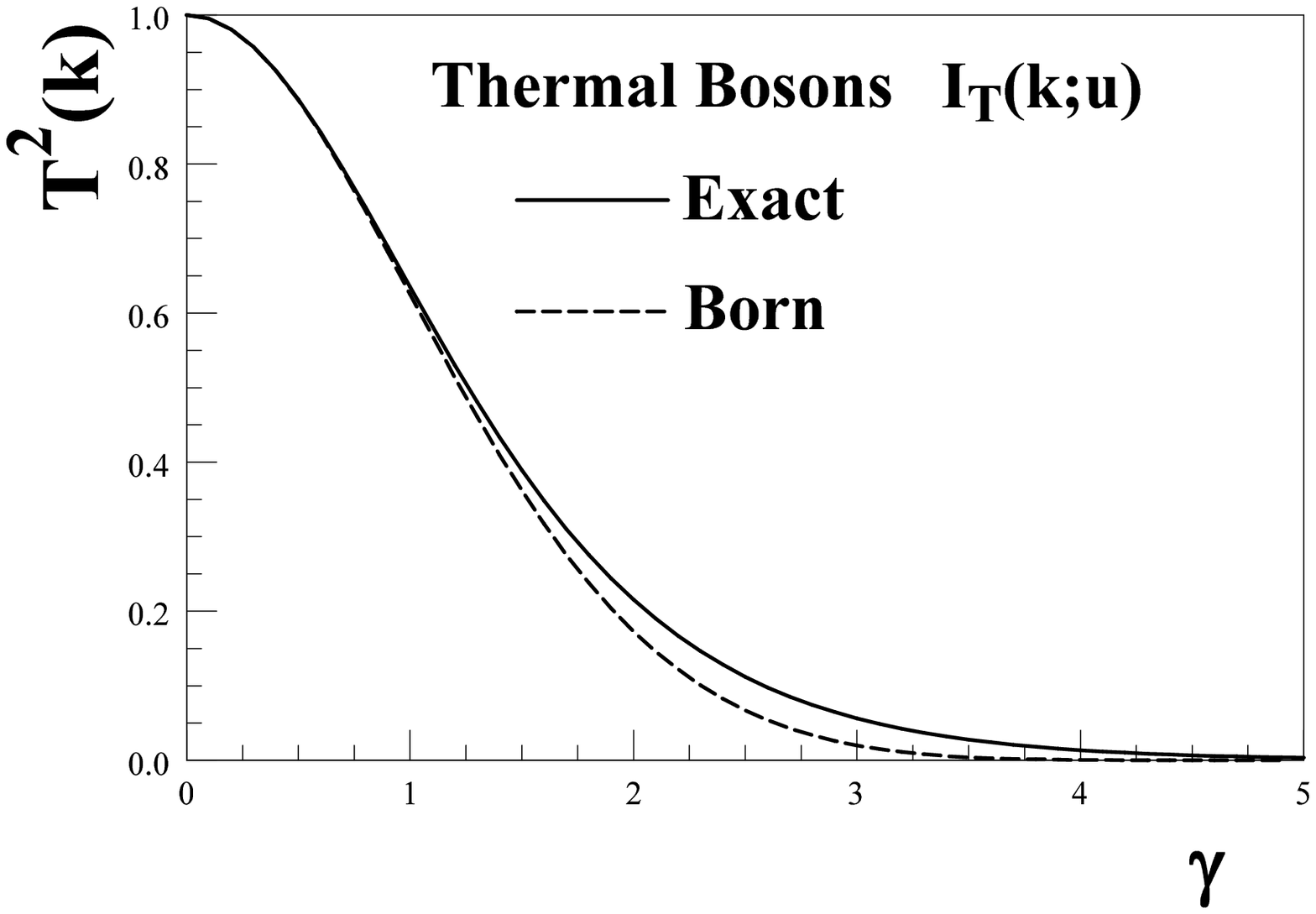}
\includegraphics[height=2in,width=3in,keepaspectratio=true]{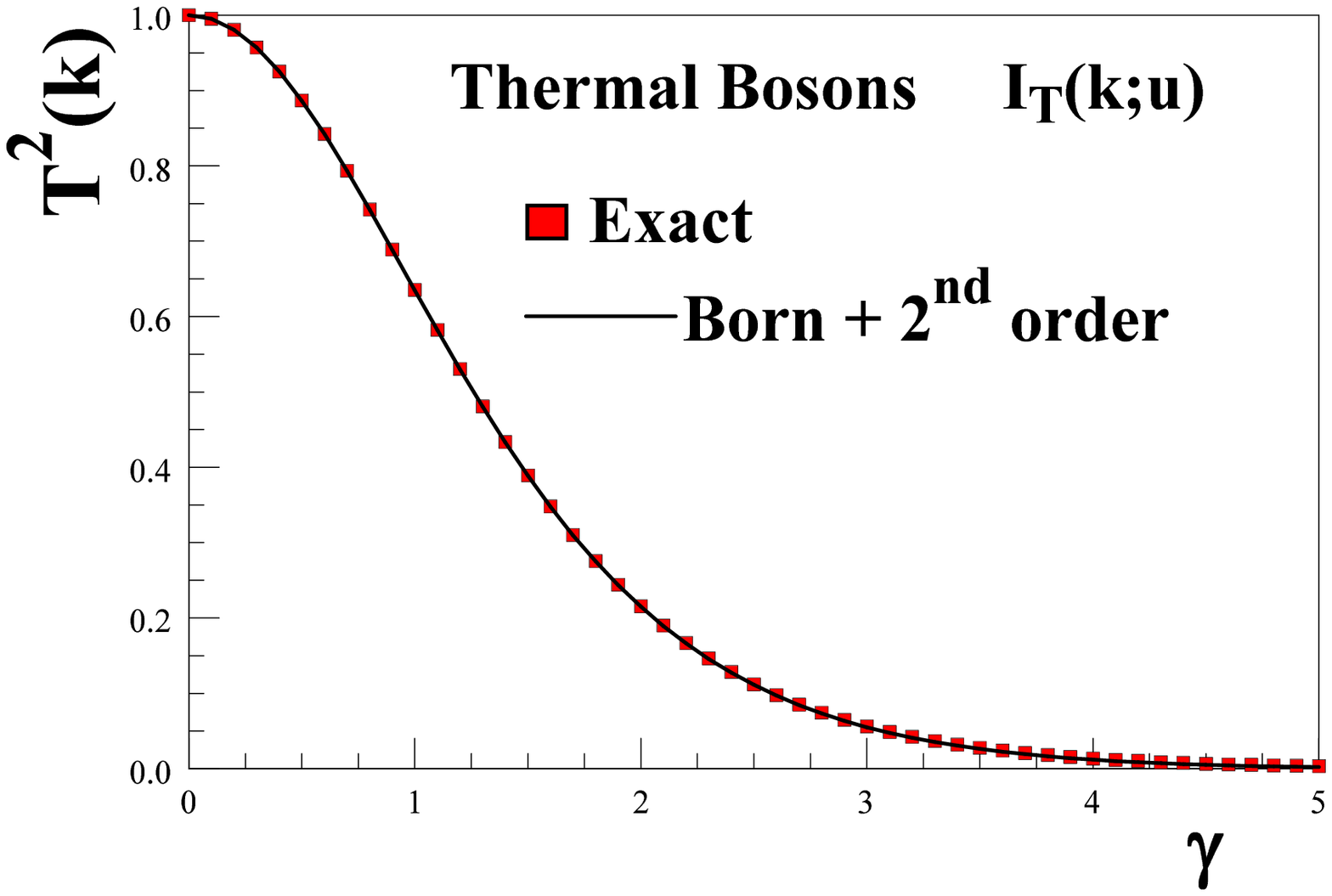}
\caption{Bosonic thermal relics.
Left panel: $ T^2(k) $ vs $ \gamma $ for the exact solution and
the Born approximation. Right panel: $ T^2(k) $ vs. $ \gamma $ for the
exact solution and the Born approximation plus second order.
Initial condition given by eq.(\ref{ITBE}) corresponding to
temperature perturbations.}
\label{fig:thermalbosonIT}
\end{center}
\end{figure}

We have studied numerically the transfer function
for both initial conditions and compared the exact numerical
solution of Gilbert's equation to the Born term and the second order
correction. Both cases are qualitatively similar. Fig.
\ref{fig:thermalbosonIT} displays the results of the analysis in
the case of the initial conditions corresponding to temperature
perturbations eq.(\ref{ITBE}). The results for Gilbert's initial
conditions eq.(\ref{IGBE}) are qualitatively the same, with a
remarkable agreement between the exact numerical solution and the
second order improved Born approximation.

\medskip

$ T(k) $ for large scales follows from eq.(\ref{Talf2}) and
$$
I_G[\alpha  \, u] \buildrel{ \alpha \; u \to 0}\over= 1 - 2 \;
\frac{\zeta(5)}{\zeta(3)} \; (\alpha \; u)^2 + {\cal O}\left([\alpha \; u]^4\right) \; ,
$$
for bosons. We find that the large scale expression for $ T(k) $
eq.(\ref{taluniv}) is {\bf also valid} in the BE case.

\medskip

The   numerical solution of eq.(\ref{gileq}) for bosons shows that
$ T(k) $ both for temperature and for Gilbert initial conditions
[eqs.(\ref{adiapert}) and  (\ref{gilini})] decreases as a power for $ \gamma > 6 $:
\bea \label{fitbo}
 T(k) \simeq \left(\frac{\gamma_B}{\gamma}\right)^{x_B} \quad , \quad
&& x_B \simeq 4.8 , \;  \gamma_B \simeq 2.9 \; \; {\rm for~ temperature ~ initial ~ conditions} \; ,
\cr \cr  
&& x_B \simeq 4.6 , \;  \gamma_B \simeq 3.2 \;  \; {\rm for~ Gilbert ~ initial ~ conditions} \; .
\eea
Again the scale of suppression of $ T(k) $ {\bf increases} with increasing $ k $,   from $ \gamma_0
\simeq 1.62 \ldots $ for small $ k $ to $ \gamma_{B} \simeq 2.9 - 3.2 $
for $ \gamma > 6 $. However, these scales are all of the same order $\sim \mathrm{O}(1)$, evidence of
a \emph{unique characteristic scale} $\gamma_{char} \sim \mathrm{O}(1)$. Namely, just as in the previous cases, in terms
of the wavevector $k$ the relevant scale of suppression is $k_{fs}(t_{eq})$.

\section{Small scales: statistics and memory of gravitational
clustering}

The difference in statistics, namely the distribution function of
the decoupled particles, enters in the initial condition and in the
non-local kernel of Gilbert's equation (\ref{gil2}). The
initial conditions are determined from the evolution of perturbations
during the cosmological stages prior to matter domination and their
dependence on the
distribution function may be different from the ones studied in the
previous sections, while the kernel in eq.(\ref{gil2}) is independent of the
initial conditions.

\medskip

The study above revealed a remarkable
difference arising from the different distribution functions. We
have established that the free streaming solution described by
$ I[\alpha \, u] $ has very different asymptotics, falling off exponentially with
$ \alpha \; u $ in the case of
Maxwell-Boltzmann and as power laws in the Fermi-Dirac and
Bose-Einstein cases. Furthermore, the fall-off is much faster for
fermions than for bosons. Since the maximum value of $ u $ is
$ u =1 $ the asymptotic behavior in $ \gamma \; u $ actually describes
the region $ \gamma \gg 1 $ or scales much smaller than the free
streaming scale.

\medskip

We have also highlighted that the different statistics lead
to an important difference in the contributions from the kernel $ K(u - u') $:
whereas Maxwell-Boltzmann statistics leads to a short range
memory that falls off exponentially in $ (u - u') $
eq.(\ref{KMB}), both Fermi-Dirac and Bose-Einstein
lead to long-range memory that falls off with
a power law in $ (u - u') $ [eqs.(\ref{Kfasi}) and (\ref{Kbasi})],
the smallest power, namely the slowest fall-off
corresponds to Bose-Einstein statistics as a consequence of the
infrared enhancement (large population at small momentum).

\medskip

\begin{figure}[ht]
\begin{center}
\begin{turn}{-90}
\includegraphics[height=13cm,width=8cm]{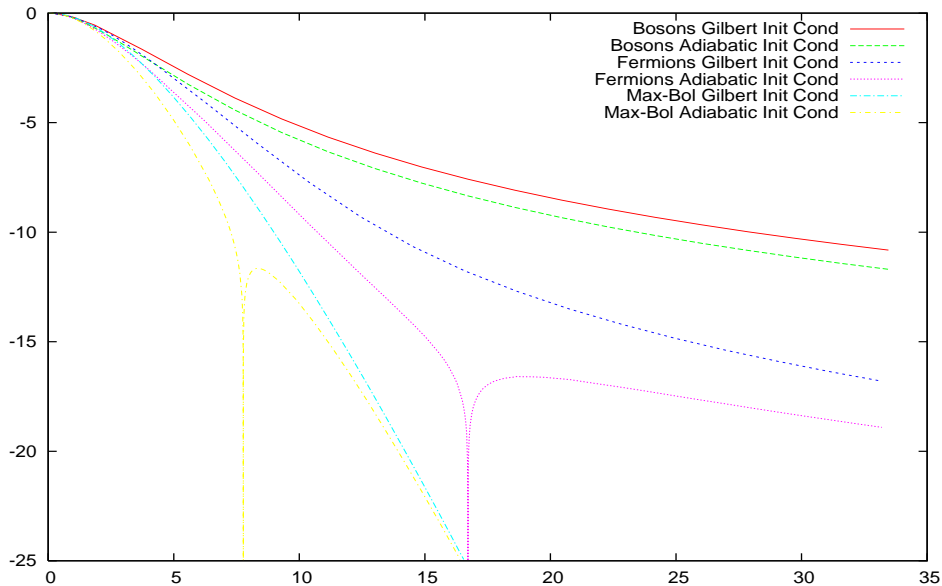}
\end{turn}
\caption{Comparison of exact numerical solutions of eq.(\ref{gil2})
and (\ref{Tofk2}) for $ \ln |T(k)| $ vs.
$ \gamma $ for Maxwell-Boltzmann particles, fermion and boson thermal relics with
Gilbert and temperature initial conditions eqs.(\ref{adiapert}) and (\ref{gilini}).  }
\label{fig:gamagrande}
\end{center}
\end{figure}

Furthermore, we have shown in the previous sections that the
influence of the kernel $ K(u-u') $ (the second order correction)
becomes important at small scales ($ \gamma > 1 $).
The long-range of the kernel keeps memory of the early stages
of the evolution when the gravitational perturbation has its largest
amplitude, therefore the long-range nature of the kernel leads to an
\emph{enhancement} of the transfer function and the power spectrum
at small scales as depicted in fig. \ref{fig:gamagrande}.

\medskip

We can study the small scale behavior in more detail by focusing on
the time scales during which the gravitational potential varies
slowly and a Markovian approximation, such as that discussed in
section \ref{MarkoMB} is reliable. As discussed above and
explicitly shown by the numerical study, for $ \gamma \gg 1 $ the
gravitational potential decays very rapidly over a time scale $ u\sim
1/\gamma $, then after evolves slowly until $ u\sim 1 $. In order to
gain deeper understanding of the dynamics during this time scale it
is convenient to implement a Markovian approximation directly  in
Gilbert's equation (\ref{gil2}). Upon changing variables in the
integrand in eq.(\ref{gil2}) $ \alpha(u-u') \equiv z $, eq.(\ref{gil2})
becomes
\be \label{gilnew}
\Phi(k,u) - \frac6{\alpha^2} \; (1-u)^2 \int^{\alpha
u}_0 \Pi[z] \; \frac{\Phi(k,u-\frac{z}{\alpha})}{\big[1-
(u-\frac{z}{\alpha})\big]^4} \;
dz = (1-u)^2  \; I[\alpha \, u] \; ,
\ee
where the kernel   $ \Pi[z] $ is given by eq.(\ref{sigma}).

For the Maxwell-Boltzmann case [see eq.(\ref{gileq})] it is more
convenient to change variables to $ \gamma(u-u') \equiv z $.

The kernel $\Pi[z]$ vanishes at $ z =0 $, is sharply peaked near $ z\sim 1 $ and falls off
exponentially for Maxwell-Boltzmann (see eq.(\ref{gileq})), as $ 1/z^3 $ for
Fermi-Dirac and as $ 1/z $ for Bose-Einstein.

For $ u \gg 1/\alpha $ and $ (1-u) \gg 1/\alpha $ the terms that multiply $ \Pi[z] $
may be expanded in a power series expansion in $ z/\alpha $.
This is the Markovian approximation, in which we obtain
\be \label{gilMarko}
\Phi(k,u) \; \Gamma[k;u] + \frac6{\alpha^3} \; (1-u)^2  \; \frac{d}{du}
\Bigg[\frac{\Phi(k,u)}{(1-u)^4}\Bigg]
\int^{\alpha u}_0 \Pi[z] \; z \; dz
\left[1+{\cal O}\left(\frac{z^2}{\alpha^2}\right)\right]=(1-u)^2
\; I[\alpha \, u] \; .
\ee
The first order equation (\ref{gilMarko}) has a simple exponential
solution that determines  the decaying behavior of the gravitational
perturbation, where the effective decay rate of gravitational
perturbations $ \Gamma[k;u] $ is given by
\be \label{Gaker}
\Gamma[k;u]=1-\frac6{\alpha^2 \; (1-u)^2} \int_0^{\alpha u} \Pi[z] \; dz
\equiv 1-\frac{K^2_{\Phi}(t)}{k^2} \; .
\ee
The wave-vectors leading to the smaller positive values of the decay rate
$ \Gamma $ are the ones that decay the least and for which the power is
the largest.
This is akin to  the criterion that determines the Jeans wave vector
in a fluid with gravitational perturbations: the Jeans wave vector
separates the stable acoustic oscillations from the unstable,
growing modes corresponding to gravitational collapse. This is
also the criterion that determines the free streaming wave-vector
in Minkowski space-time \cite{freestream}.

For Maxwell-Boltzmann and Fermi-Dirac distribution functions the upper limit in the
integrals can be safely taken to infinity for $ \alpha \; u \gg 1 $, with the result
\bea \label{integz}
\int_0^\infty \Pi[z]  \; dz & = &  \int_0^\infty \tilde{f}_0(y) \; dy=
\Big\langle \frac1{y^2}\Big\rangle  \\\label{Con}
\int_0^\infty z \;  \Pi[z] \;  dz & = & \mathcal{A} \; .
\eea
where the (finite) constant $\mathcal{A}$ depends on the distribution function and the angular brackets stand for the average with $ \tilde{f}_0 $.
In these cases we find for $ K^2_{\Phi}(t) $ [see eqs.(\ref{hom}) and
(\ref{alfa})]
\be \label{KFSu}
K^2_{\Phi}(t) = \frac32 \; H^2_{0M}  \;  \frac{a_{eq}}{(1-u)^2} \;
\Big(\frac{m}{T_{d,0}}\Big)^2~\Big\langle \frac1{y^2}\Big\rangle
   = 4 \, \pi  \; G  \; \rho_{0M} \; \Big\langle
   \frac1{\vec{V}^2}\Big\rangle \; a(t) = K^2_{\Phi}(0) \; a(t) \; .
\ee
Remarkably, $ K_{\Phi}(t) $ coincides with the free streaming wave-vector
found in Minkowski space-time \cite{freestream},
\be \label{KFSss}
K_{\Phi}(0) =  \Bigg[4 \, \pi  \; G  \; \rho_{0M} \; \Big\langle
\frac1{\vec{V}^2}\Big\rangle \Bigg]^\frac12 =
0.563\, \Bigg[\Big\langle \frac1{y^2}\Big\rangle\Bigg]^\frac{1}{2} \; \Big( \frac{g_d}2 \Big)^\frac13 \;
\frac{m}{\mathrm{keV}} \; [\mathrm{kpc}]^{-1} \; .
\ee
Thus, at small scales the Minkowski result is obtained, consistently with
the expectation that at short wavelengths an adiabatic approximation to
the expansion is reliable \cite{freestream,ringwald}.

The full transfer function $ T(k) $ cannot be obtained from the Markovian
approximation alone and the full study presented in the previous
sections is necessary. However, it becomes clear that
$ k_{fs}(t_{eq}) $ is not the \emph{only} relevant scale and that
there is a further scale $ K_{\Phi}(t_{eq}) $ eq.(\ref{KFSss}).
The ratio of these two expressions is
\be \label{ratiokfs}
\frac{K^2_{\Phi}(0)}{k^2_{fs}(0)} = \langle \vec{V}^2 \rangle \;
\Big\langle \frac1{\vec{V}^2}\Big\rangle
= {\int^\infty_0 y^4 \; \tilde{f}_0(y) \; dy}{\int^\infty_0
   \tilde{f}_0(y)\; dy}= \Bigg\{\begin{array}{l}
                      3~~\mathrm{Maxwell-Boltzmann} \\
                      4.9742\ldots~~\mathrm{Fermi-Dirac}
                    \end{array}    \; .
\ee
Defining the small scale length \emph{today}
\be \label{Lambda}
\Lambda_{\Phi}(t_{eq}) = \frac{2 \, \pi}{K_{\Phi}(t_{eq})}
\ee
we find for WIMPs,
\be \label{Lamwim}
\Lambda_{\Phi}(t_{eq}) \sim 0.5 \; \mathrm{pc} \; \Big(\frac2{g_d}\Big)^\frac13  \;
\Big(\frac{100\,\mathrm{GeV}}{m} \; \Big)^\frac12 \;
\Big(\frac{10 \,\mathrm{MeV}}{T_d}\Big)^\frac12 \; ,
\ee
and for thermal fermions,
\be \label{LamFD}
\Lambda_{\Phi}(t_{eq})=  939 \; \mathrm{kpc} \;
\Big(\frac2{g_d} \Big)^\frac13 \; \frac{\mathrm{keV}}{m} \quad {\rm FD} \quad
  \; .
\ee

   The results above are valid for both Maxwell-Boltzmann and
   Fermi-Dirac statistics, in these cases the kernel $ \Pi[z] $ falls
   off fast enough to make its integral finite. This is \emph{not}
   the case for Bose-Einstein statistics for particles that
   decoupled in LTE while ultrarelativistic. In this case the
   asymptotic behavior of $ \Pi[z] \propto 1/z $ leads to an
   \emph{infrared} enhancement as a consequence of the long-range of
   the kernel. In this case we keep the upper limit in eq.(\ref{Gaker})
and carry out the integral in $ z $, leading to \cite{DRG}
\be  \label{intpi}
\int_0^{\gamma u}\Pi[z]\,dz = \frac1{2 \, \zeta(3)}
\int_0^\infty \frac{dy}{e^y -1} \; \Big[1-\cos(\gamma \, u\,y)\Big]
\buildrel{\gamma \; u \to \infty}\over=
\frac{\ln[\gamma \; u \; e^{\cal C}]}{2 \; \zeta(3)} \; .
\ee
where $ {\cal C} = 0.577216\ldots $ is the Euler-Mascheroni constant
as well as in the coefficient of the derivative term in
eq.(\ref{gilMarko}), leading to
\be \label{zpiz}
\int_0^{\gamma \; u}z \; \Pi[z] \; dz
\buildrel{\gamma \; u \to \infty}\over =
\frac{\gamma \; u}{2 \, \zeta(3)} \; .
\ee
This infrared enhancement reflects the infrared divergence of the
   free streaming wavevector in Minkowski space-time found in
   ref.\cite{freestream}. The argument of the logarithm in
   eq.(\ref{intpi}) clearly reveals that it is the cosmological
   expansion that yields an infrared cutoff. Taking $ u\sim 1 $
   [neglecting terms of $ \mathcal{O}(1/\gamma) $] we find a
   sliding wavevector at small scales for the
   Bose-Einstein case, namely
\be \label{KFSssBE}
K_{\Phi}(t_{eq}) \simeq 0.00424\ldots \; \sqrt{\ln\gamma + {\cal C} } \;
\Big( \frac{g_d}{2}\Big)^\frac13 \; \frac{m}{\mathrm{keV}} \;
[\mathrm{kpc}]^{-1} \quad {\rm BE} \quad \; .
\ee
The larger $ K_{\Phi}(t_{eq}) $ leads to shorter free streaming lengths and
to more power at small scales because free streaming smoothes out on
shorter scales. Therefore, at \emph{small scales} BE particles
that decoupled while relativistic behave as CDM. This conclusion
is in agreement with the results in refs. \cite{coldmatter,freestream} and is borne
out in the numerical solution displayed in fig. (\ref{fig:gamagrande}).

In all cases  $ K_{\Phi}(t_{eq}) $ controls the decay of the gravitational
fluctuations $ \Phi(k,u) $ for small scales and long times. The logarithmic behavior
and the consequent increase of $K_{\Phi}$ yields an explanation of the enhancement at
small scales over the FD and MB cases depicted in fig. (\ref{fig:gamagrande}).


\medskip

{\bf Comparing statistics and initial conditions:}

The analysis presented above clearly indicates the differences arising from statistics
and initial conditions. It proves convenient to compare the results of the numerical solution
of Gilbert's equations for all cases considered parametrized by the dimensionless ratio $\gamma$.
This comparison is depicted in fig. \ref{fig:gamagrande} for a wide range of $\gamma$. The
differences from statistics and initial conditions are clearly displayed in this figure, and are
in complete agreement with the analysis presented above.

For thermal relics that decoupled while relativistic, either FD or BE, the range of
scales relevant for structure formation in which the linearized approximation is reliable
corresponds to $0 < \gamma \leq 6$, whereas for WIMPs, the relevant range of scales corresponds
to   $\gamma \leq 10^{-5}$. We have confirmed that in the range of cosmological relevance for
all species, the second order approximation (\ref{Tint2ndord}) for $T(k)$ is \emph{very  accurate} and
indistinguishable from the exact numerical solution in the range $0 < \gamma \leq 6$.

As shown in fig. \ref{fig:gamagrande} $T(k)$ features zeroes  for MB and FD statistics for temperature
initial conditions. However the value of $\gamma$ for FD at which $T(k)$ vanishes corresponds
to a sub-galactic scale, and for MB  it corresponds to a sub-parsec scale, in either case well
outside the regime of reliability of the linearized approximation, hence these features are not
relevant for structure formation. Nevertheless, this figure
highlights the main aspects discussed above: for a \emph{fixed} value of $\gamma$  BE statistics favors the small momentum region,
leads to a longer range memory kernel that falls off with a power law and yields the largest $T(k)$ for a fixed value of $\gamma$, followed by FD statistics with a (slower) power law fall off and finally
the MB distribution with an exponential fall off of the memory kernel and the smallest $T(k)$ for
fixed $\gamma$.

\section{Conclusions}

In this article we studied the evolution of gravitational and DM
density perturbations from the collisionless Boltzmann-Vlasov during matter domination,
and obtained an \emph{exact} expression for the transfer function $T(k)$ for \emph{arbitrary
distribution function of the decoupled DM particle and initial conditions}.

We have transformed the non-relativistic Boltzmann-Vlasov equation into an integro-differential
equation which features a non-local kernel that describes the memory of gravitational clustering and yields corrections to the fluid description. This formulation lends itself to a systematic Fredholm expansion for the evolution of DM density and gravitational perturbations and  $T(k)$, and makes explicit the
influence of the distribution function on $T(k)$.

Distribution functions that favor the small momentum region lead to \emph{longer range} kernels,
a persistence of the memory of the initial conditions and gravitational clustering resulting in
an \emph{enhancement} of $T(k)$ at \emph{small scales}.

The natural scale of suppression of $T(k)$ is determined by  the free-streaming wave vector at matter-radiation equality,   $k_{fs}(t_{eq}) = \sqrt{6}/l_{fs}(0)$
where the comoving free streaming wavevector is \be k_{fs}(t) =
\left[\frac{4\pi \rho_{0M} a(t)}{\langle \left(\frac{\vec{p}}{m}
\right)^2 \rangle} \right]^\frac{1}{2}\, ,\label{kfs1} \ee   the
angular brackets refer to the average with the distribution function
of the decoupled particle, and $l_{fs}(0)$ is the comoving free
streaming distance traveled by the decoupled particle from the time
of matter-radiation equality $t_{eq}$ until today.

We find \be k_{fs}(t_{eq}) =
\frac{k_{fs}(0)}{\sqrt{1+z_{eq}}} \ee with \bea k_{fs}(0) = \left\{
\begin{array}{l}
   \frac{0.325}{\big[10^{-3}\mathrm{pc}\big]}\, \Big( \frac{g_d}{2}
\Big)^\frac{1}{3} \Big(\frac{m}{100\,\mathrm{GeV}}
\Big)^\frac{1}{2}\Big(\frac{T_d}{10 \,\mathrm{MeV}}
\Big)^\frac{1}{2} ~~\mathrm{WIMPs} \\
     0.157\, \Big(
\frac{g_d}{2}\Big)^\frac{1}{3}\Big(\frac{m}{\mathrm{keV}} \Big)\,
[\mathrm{kpc}]^{-1} ~~\mathrm{FD~thermal~relic} \\
     0.175\, \Big(
\frac{g_d}{2}\Big)^\frac{1}{3}\Big(\frac{m}{\mathrm{keV}} \Big)\,
[\mathrm{kpc}]^{-1} ~~\mathrm{BE~thermal~relic}
         \end{array} \right.\eea where $g_d$ is the number of relativistic species at
         decoupling.

We provided a detailed numerical study for thermal relics, WIMPs and fermionic and
bosonic particles that decoupled while relativistically.
The result of the numerical study of $T(k)$  as a
function of $ \gamma = \sqrt{2}k/k_{fs}(t_{eq})$ for the three cases and different initial
conditions is displayed in fig.  \ref{fig:gamagrande}.

This study reveals that the first \emph{two} terms in the Fredholm expansion
yield a remarkable accurate approximation to $T(k)$ in the range of scales of cosmological
relevance for structure formation. In all cases considered we find that the exact solution differs from the second order approximation in
less than $5\%$ in the range $0<\gamma\lesssim 6$ corresponding to scales down to  a fraction of the free-streaming length
and much less than this on large scales for $0 < \gamma \lesssim 1$.

\medskip

A simple and accurate approximation to $T(k)$ is given by eqn. (\ref{Tint2ndord}). The
second term in this expression includes corrections beyond the fluid approximation and
higher order moments in the Boltzmann hierarchy and describes the memory of initial conditions
and gravitational clustering. It explicitly depends on the distribution function of the
decoupled particle. Non-local kernels with longer range lead to an \emph{enhancement} of
$T(k)$ at small scales. FD and BE thermal relics feature kernels with longer range than WIMPs,
with BE statistics (for relics that decoupled relativistically) leading to longer range and
more power at small scales.

This behavior is   clearly exhibited   in fig. \ref{fig:gamagrande}.

For long scales $k \ll k_{fs}(t_{eq})$ we find the behavior \be T(k) = 1- \Bigg( \frac{\gamma}{\gamma_0}\Bigg)^2+\cdots \ee where  $\gamma_0 $ is given by (\ref{gamacero}) for the
initial conditions considered here independent of the statistics for thermal relics.

We provide   fits of $T(k)$, within a wide range of (small) scales for $k>k_{fs}(t_{eq})$
in the intervals where it exhibits a simple
power-like or exponential behavior for MB, FD and BE statistics: see
 eqs.(\ref{fitMB}), (\ref{fitfg}), (\ref{feradia}), (\ref{fa2})
and (\ref{fitbo}).

Although we do not attempt to provide a global fit with \emph{one} function, it is clear that
in the small scale region, the functional forms of $T(k)$ found in this article by exact
numerical solution of the Boltzmann-Vlasov equation (in Gilbert's form) are very different from
the  fits by Bardeen \emph{et.al.}\cite{BBKS} often quoted in the literature.

\medskip

An important consequence of this study is a distinct imprint of
the particle statistics and
its distribution function on the transfer function at \emph{small scales}, $ \lambda \ll l_{fs}(0) $:
a distribution function that enhances the \emph{small} momentum region yields a
\emph{longer range memory} of the initial conditions and gravitational clustering
and an \emph{enhancement} of the transfer function at small scales as depicted
in fig. \ref{fig:gamagrande}.
This result may prove important in the elucidation of the small scale structure of DM
halos and perhaps lead to an explanation of the filamentary structures found in numerical
simulations in ref.\cite{gao}.

The tools provided in this article to study the transfer function for
arbitrary distribution functions and general initial conditions, in particular
the simple and remarkably accurate approximation to $T(k)$ given by eqn. (\ref{Tint2ndord})
allows a systematic and robust assessment of different DM candidates.

\acknowledgements{D.B. thanks Andrew Zentner and Alex Kusenko for stimulating
discussions and insight. He acknowledges support
from the U.S. National Science Foundation through grant No:  PHY-0553418.}

\end{document}